\pdfoutput=1

\documentclass[aps,pra, amsmath,superscriptaddress, longbibliography,twocolumn, 10pt]{revtex4-1}
\usepackage{siunitx}
\usepackage[pdftex]{graphicx}
\usepackage{bm}
\usepackage{setspace}

\begin{document}
\setstretch{1.05}

\title{Valley-Polarized Exciton-Polaritons in a Monolayer Semiconductor}
\author{Yen-Jung Chen}
\affiliation{Department of Physics and Astronomy, Northwestern University, Evanston, Illinois 60208, USA}
\author{Jeffrey D. Cain}
\affiliation{Department of Materials Science and Engineering, Northwestern University, Evanston, Illinois 60208, USA}
\affiliation{International Institute for Nanotechnology, Northwestern University, Evanston, Illinois 60208, USA}
\author{Teodor K. Stanev}
\affiliation{Department of Physics and Astronomy, Northwestern University, Evanston, Illinois 60208, USA}
\author{Vinayak P. Dravid}
\affiliation{Department of Materials Science and Engineering, Northwestern University, Evanston, Illinois 60208, USA}
\affiliation{International Institute for Nanotechnology, Northwestern University, Evanston, Illinois 60208, USA}
\author{Nathaniel P. Stern}\email{n-stern@northwestern.edu}
\affiliation{Department of Physics and Astronomy, Northwestern University, Evanston, Illinois 60208, USA}

\begin{abstract}
\vspace{1em}
Single layers of transition metal dichalcogenides are two-dimensional direct bandgap semiconductors with degenerate, but inequivalent, `valleys' in the electronic structure that can be selectively excited by polarized light. Coherent superpositions of light and matter, exciton-polaritons, have been observed when these materials are strongly coupled to photons, but these hybrid quasiparticles do not harness the valley-sensitive excitations of monolayer transition metal dichalcogenides. Here, we demonstrate evidence for valley polarized exciton-polaritons in monolayers of MoS$_2$ embedded in a dielectric microcavity. Unlike traditional microcavity exciton-polaritons, these light-matter quasiparticles emit polarized light with spectral Rabi splitting. The interplay of cavity-modified exciton dynamics and intervalley relaxation in the high-cooperativity regime causes valley polarized exciton-polaritons to persist to room temperature, distinct from the vanishing polarization in bare monolayers. Achieving polarization-sensitive polaritonic devices operating at room temperature presents a pathway for manipulating novel valley degrees of freedom in coherent states of light and matter.
\vspace{2em}
\end{abstract}

\maketitle
\vspace{2em}

Confining photons to small volumes, such as between two mirrors, can enhance light-matter interactions and lead to coherent superpositions of light and matter, or polaritons~\cite{Purcell1946, Haroche1989}. This framework of cavity quantum electrodynamics (QED) is the foundation for strong, coherent interactions with light in both atoms~\cite{Kimble1998} and two-level solid state systems~\cite{Reithmaier2004, Yoshie2004, Wallraff2004}. Translating this approach to bosonic exciton-polariton quasiparticles in semiconductors~\cite{Weisbuch1992, Houdre1994} enables new many-body coherent phenomena such as Bose-Einstein condensation in the solid-state~\cite{deng2010exciton}, with potential opto-electronic applications in `polaritonics'~\cite{amo2010exciton, Sanvitto2016}. Although the cavity QED parameter regimes vary greatly across these implementations, they generally share the spectral selectivity of high-quality resonators.

More recently, polarization selectivity has brought a new dimension to cavity QED.  For example, accessing specific sublevels using polarized light can merge time-reversal symmetry breaking into chiral photonics~\cite{Lenferink2014, Sayrin2015, Sollner2015}. Although individual atoms allow strong coupling in circularly-polarized cavity QED~\cite{Junge2013}, the degenerate valley-specific excitons of transition metal dichalcogenides (TMDs) are a model material system for achieving polarization selectivity with bosonic exciton-polaritons~(Fig.~\ref{figure1}a).

TMDs are layered two-dimensional (2D) semiconductors that have a direct band gap in their monolayer form~\cite{mak2010atomically,splendiani2010emerging}. Inversion asymmetry in TMDs leads to strong spin-orbit splitting at the direct bandgap of the $K$ and $K^\prime$ valleys. Monolayer TMDs such as MoS$_2$ or WSe$_2$ can therefore support two different classes of energy degenerate excitons, identical in most properties, but with opposite Berry curvature and distinct response to light of opposite helicity depending on their valley pseudospin index~\cite{xiao2012coupled} (Fig.~\ref{figure1}b). The selectivity of valley phenomena has been exploited for polarization-dependent opto-electronics~\cite{mak2012control, zeng2012valley,cao2012valley,kioseoglou2012valley, wang2012electronics, sallen2012robust}, correlated spin-electron motion~\cite{Mak2014}, and proposed mechanisms for coherent information processing~\cite{Rohling2012, Behnia2012, Jones2013, Wang2016}. Controlling the dynamics of valley-polarized excitons with polarization-sensitive cavity QED is an exciting possibility not yet understood.

Monolayer TMD materials can be interfaced with photonic cavities~\cite{gan2013controlling,wu2014control,schwarz2014two, Wei2015} to enhance radiative coupling. As with excitons in semiconductor quantum wells, strongly coupled 2D exciton-polaritons have been observed by embedding monolayer TMDs in planar microcavities (MCs)~\cite{liu2015strong}. The tightly bound excitons and large oscillator strengths in TMDs allow these quasiparticles to persist at room temperature~\cite{liu2015strong, dufferwiel2015exciton, Flatten2016, Lundt2017}. Marrying the valley structure of excitons in TMD monolayers with cavity QED will extend polaritonics to a new regime in which two species of degenerate exciton-polariton quasiparticles can spatially co-exist in separate regions of momentum space and can be selectively excited by helical cavity fields.  Understanding this polarization-sensitive cavity QED regime would make possible new explorations of coherent optical phenomena in 2D semiconductors and their layered heterostructures~\cite{Low2016, Basov2016}.

Here, we report the observation of valley-polarized exciton-polaritons in monolayer MoS$_2$ embedded in a planar microcavity~(Fig.~\ref{figure1}a). These distinct degenerate light-matter quasiparticles are selectively coupled to circularly-polarized cavity modes. In contrast to excitons in bare monolayers, the MC exciton-polaritons preserve valley polarization at room temperature, which is explained by the competition between the exciton valley decay time and the cavity-enhanced total decay rate of the exciton-polaritons in the cavity QED system. The ability to manipulate pseudospin dynamics in hybrid optical devices with 2D materials opens new opportunities for valley-sensitive 2D material photonics with engineered chiral light-matter interactions.

\section*{Results}

\begin{figure}[tbp]
\includegraphics[scale=1]{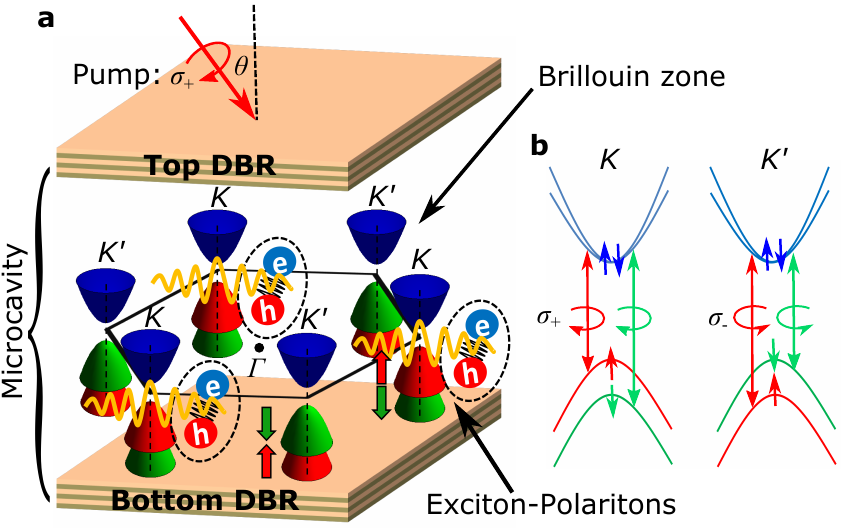}
\caption{\textbf{Valley-polarized exciton-polaritons in MC-MoS$_2$.} \textbf{a}, Schematic of valley-polarized exciton-polaritons in a MC. Monolayer MoS$_2$ is represented in reciprocal space. When the MC is pumped by circularly polarized light ($\sigma_{+}$) at angle $\theta$, strongly-coupled exciton-polaritons are created primarily in the $K$ valleys. \textbf{b}, Valley and spin optical transitions of MoS$_2$.} \label{figure1}
\end{figure}

The monolayer MoS$_2$ microcavity (MC-MoS$_2$) samples consist of a single layer of monolayer MoS$_2$ embedded in a distributed Bragg reflector (DBR) microcavity (MC), as illustrated in Fig.~\ref{figure1}a. The DBRs are made from alternating silicon dioxide and silicon nitride layers grown on top of a Si substrate.
Separate MCs were fabricated both with the cavity on resonance with and detuned from the MoS$_2$ $A$-exciton for use in different experiment configurations, as we will describe. The monolayer flakes of MoS$_2$ are synthesized by chemical vapor deposition and transferred on top of the bottom DBR before the top DBR is deposited so they are at the center of the inner cavity.   Both as-grown and transferred monolayer MoS$_2$ show $A$-exciton photoluminescence (PL) peak centered at 1.855~eV with an inhomogeneously broadened full width half-maximum linewidth $\Gamma_{\rm ex} =  64$~meV when collected with a microscope objective. Measured with a lens instead, the $A$-exciton PL is centered at 1.81~eV with linewidth of 90~meV due to the increased inhomogeneous broadening of the larger spot.

\begin{figure}[tbp]
\includegraphics[scale=1]{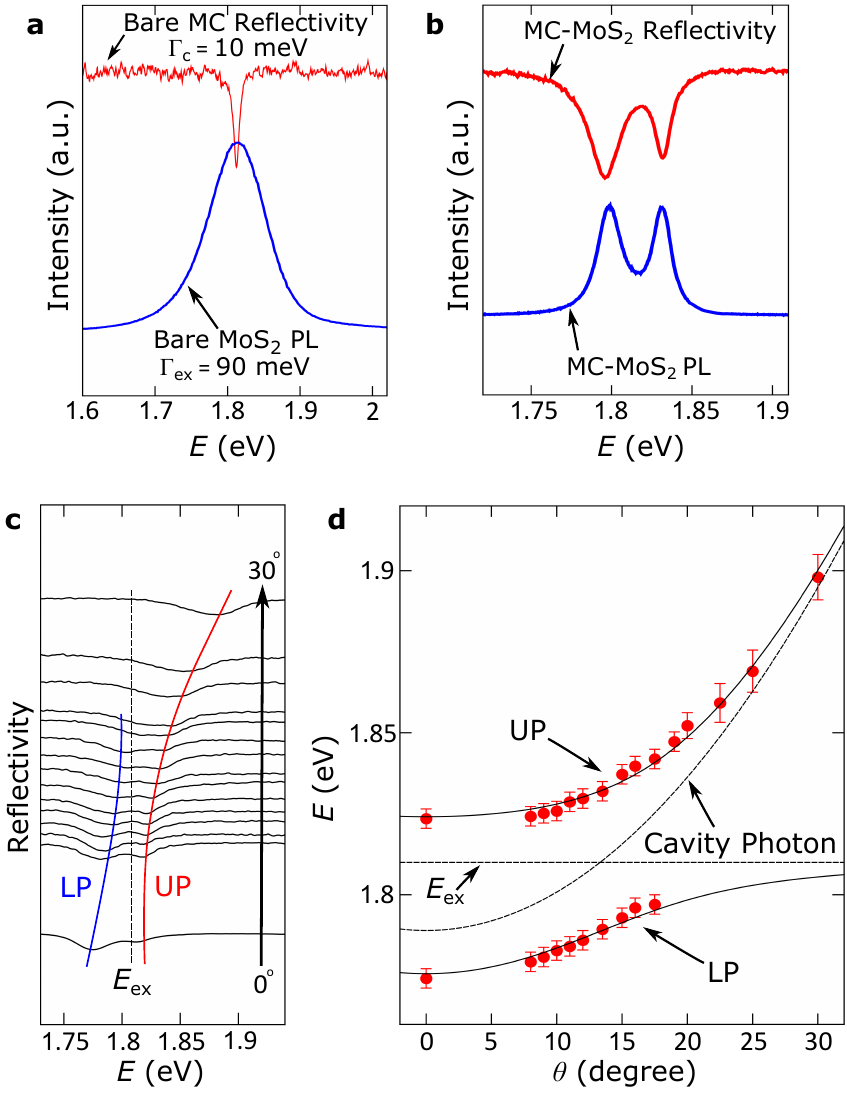}
\caption{\textbf{The dispersion of exciton-polaritons in MoS$_2$.} \textbf{a}, Reflectivity spectrum of a bare MC (top) and PL spectrum of bare monolayer MoS$_2$ (bottom). Both are inhomogeneously broadened. \textbf{b}, Reflectivity and PL spectra for MC-MoS$_2$. \textbf{c}, Angle-resolved reflectivity spectra with the MC detuned by -21~meV with respect to MoS$_2$ $A$-exciton energy. Incident angles of the white light are tuned from 0 to 30 degrees. The UP and LP energies are on the opposite sides of the exciton energy. \textbf{d}, The dispersion relation of exciton-polaritons by fitting the polariton data from the spectra in \textbf{c} with a coupled oscillator model. Anti-crossing between the UP and LP is observed with Rabi-splitting $\hbar \Omega = 39 \pm 5$~meV at $\theta = \ang{13.5} $.} \label{figure2}
\end{figure}

We first confirm the presence of strongly coupled light-matter quasiparticles in the MC-MoS$_2$ by observing Rabi splitting in both reflectivity and PL spectra. In regions of the MC where there is no monolayer MoS$_2$, the reflectivity spectrum shows only a single dip with width $\Gamma_{\rm c} \approx 10$ meV, which is somewhat broadened by the non-zero collection angle from the cavity~(Fig.~\ref{figure2}a top). In regions with monolayer MoS$_2$, the MC reflectivity (PL) spectrum shows two dips (peaks) (Fig.~\ref{figure2}b). These resonance features are signatures of the upper polariton (UP) and lower polariton (LP) quasiparticle eigenstates in the MC.

Exciton-polariton behavior at room temperature is verified from an anti-crossing in the energy dispersion obtained from angle-resolved white light reflectivity (Fig.~\ref{figure2}c). As the angle $\theta$ is swept, the cavity resonance changes energy and detunes from the exciton absorption. For this measurement, the MC is designed to be red detuned from the MoS$_2$ exciton so that the modes are resonant at a non-zero angle ($\theta \approx \ang{13.5} $) and the modified MC-MoS$_2$ dispersion is clearly visible.

For the MC-MoS$_2$ sample shown in Fig.~\ref{figure2}d, for $\theta < \ang{13.5}  $, the UP energy is nearly constant and the LP energy blue-shifts with angle. For $\theta >  \ang{13.5} $, the UP and LP behavior reverses, but the dispersion curves do not touch. This anti-crossing between the UP and LP branches is a feature of the strong light-matter coupling in which the coherent exchange rate dominates the incoherent losses. Under these conditions, a resonant excitation creates exciton-polariton quasiparticles that are nearly equal superpositions of material exciton and cavity photon. The Rabi splitting, $\hbar \Omega$, between the UP and LP on resonance is $39 \pm 5$ meV~(Fig.~\ref{figure2}d). The interaction coupling constant, $g$, between 2D excitons and cavity photons was calculated from a two-mode coupled oscillator model~\cite{savona1995quantum} as $g = 28 \pm 2$ meV. This room temperature measurement is consistent with the value reported for strongly coupled monolayer MoS$_2$ exciton-polaritons in a MC~\cite{liu2015strong}. The dispersion and mode splitting structure verify that the MC at room temperature is populated by exciton-polaritons in a regime dominated by coherent exchange between light and matter excitations.

The polariton energies do not blue-shift at lower temperatures in contrast to excitons in bare MoS$_2$ on thermally grown SiO$_2$. This feature has been explained in previous reports as due to compressive strain at the interface from the different thermal expansion coefficients of the SiO$_2$ substrate and monolayer MoS$_2$~\cite{liu2015strong}. MoS$_2$ covered with other oxides such as HfO$_2$ and Al$_2$O$_3$ also show reduced temperature-dependent shifts in band gap~\cite{plechinger2012low}.


\begin{figure}[tbp]
\includegraphics[scale=1]{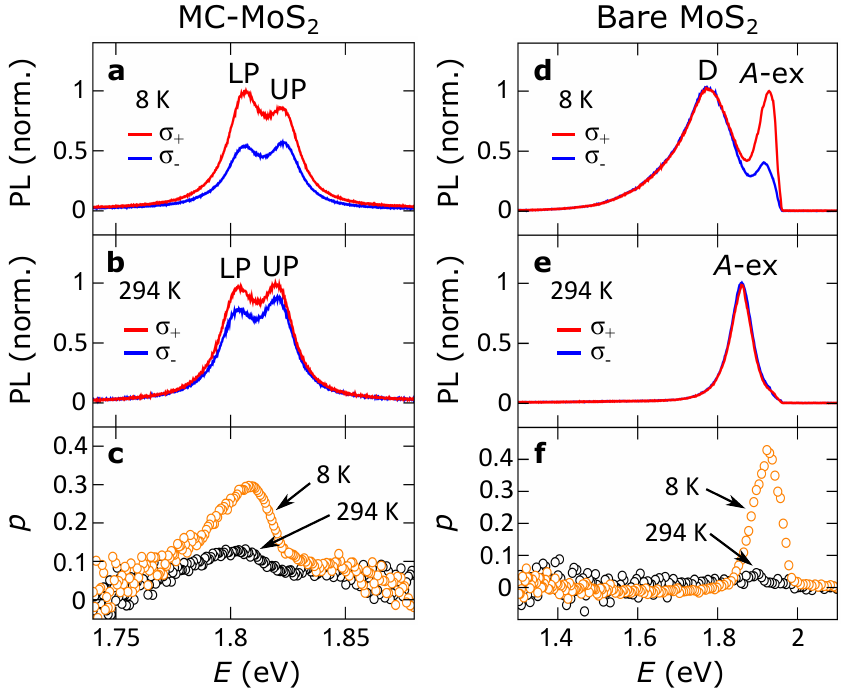}
\caption{\textbf{Valley-polarized emission.} \textbf{a}, \textbf{b}, Emission spectra of MC-MoS$_2$ pumped with $E_{\rm ph}= 1.938$~eV at 8~K and 294~K. \textbf{c}, Polarization spectrum of MC-MoS$_2$. \textbf{d}, \textbf{e}, Emission spectra of bare MoS$_2$ pumped with $E_{\rm ph}= 1.987$~eV at 8~K and 294~K. \textbf{f}, Polarization spectrum of bare MoS$_2$. The defect state is unpolarized.} \label{figure3}
\end{figure}

With the hybrid light-matter quasiparticles established in MC-MoS$_2$, the valley pseudospin polarization of the exciton-polaritons was explored using polarized photoluminescence. For this measurement, the DBR layer thicknesses were chosen so the cavity dispersion is resonant with the MoS$_2$ exciton at normal incidence. Polarized PL was first measured with near resonant pump energy $E_\textrm{ph}$ ($E_\textrm{ph} = 1.938$~eV for MC-MoS$_2$ and $E_\textrm{ph} = 1.987$~eV for bare MoS$_2$) at cryogenic temperatures (8~K) to maximize the valley polarization~\cite{mak2012control,zeng2012valley,cao2012valley,kioseoglou2012valley, sallen2012robust}. $E_\textrm{ph}$ was selected to have the same energy differences with the MC-MoS$_2$ and the bare MoS$_2$ emission at room temperature to obtain similar pumping efficiency to the valley-specific MoS$_2$ excitons. Excitons with in-plane wavevector $k_\parallel \neq 0$ are pumped in the monolayer with an off-resonant beam at an oblique angle. Momentum relaxation will populate the $k_\parallel =0$ exciton-polaritons in the MC-MoS$_2$, which will emit light collected at normal incidence. If the excitation of exciton-polaritons is valley-sensitive, then the emission from right- and left-circularly polarized cavity modes will be asymmetric. Circular polarization of $k_\parallel =0$ normal emission from Rabi-split UP and LP modes indicates that the optical properties of the MC-MoS$_2$ are determined by exciton-polariton quasiparticles with definite valley character.

The emission spectrum in Fig.~\ref{figure3}a shows the Rabi-split UP and LP resonances at 1.805~eV and 1.825~eV, respectively. Circular polarization of the cavity emission is evident. As described in the Supplementary Information, care is taken that the off-resonant circularly polarized excitation from the pump at angle $\theta = \ang{38} $ creates nearly equal $s$ and $p$ linear polarized electric field components in the DBR MC so the circular polarization of the total pumped cavity field is preserved.

To quantify the measurements, the total PL polarization $p$ is defined as
\begin{equation}
p = \frac{\left ( I_{+}-I_{-} \right )}{\left ( I_{+}+I_{-} \right )}
\label{eq:pol1}
\end{equation}
where $I_{+}$ and $I_{-}$ are the helicity-resolved PL intensities. At 8~K, both UP and LP show significant polarizations of 19$\%$ and 29.5$\%$, respectively~(Fig.~\ref{figure3}a). For bare MoS$_2$ at the same temperature (Fig.~\ref{figure3}d), we observe $p=40\%$ for the $A$-exciton, a typical MoS$_2$ polarization at cryogenic temperatures~\cite{cao2012valley,zeng2012valley,kioseoglou2012valley}. A low-temperature defect state~\cite{tongay2013defects} (labeled as $D$) is present in the bare MoS$_2$ which does not exhibit polarization, confirming the polarization arises from the valley-pumped exciton-polaritons and not measurement bias. The polarization for both UP and LP modes increases for $E_{\rm ph}$ closer to the exciton resonance, following the behavior of bare MoS$_2$~\cite{mak2012control,zeng2012valley,cao2012valley,kioseoglou2012valley, sallen2012robust} (see Supplementary Information). The emission polarization of the cavity polariton modes in the strong coupling regime confirms the presence of valley polarized exciton-polaritons in MoS$_2$ microcavities.

At a temperature of 8~K, both MC-MoS$_2$ and bare MoS$_2$ exhibit significant and comparable polarizations. Increasing to room temperature, $p$ in bare MoS$_2$ decreases close to 0$\%$ (Fig.~\ref{figure3}e, f). The MC-MoS$_2$, on the other hand, retains a non-zero polarization of 7.5$\%$ and 13$\%$ for the UP and LP. (Fig.~\ref{figure3}b, c).


\begin{figure}[tbp]
\includegraphics[scale=1]{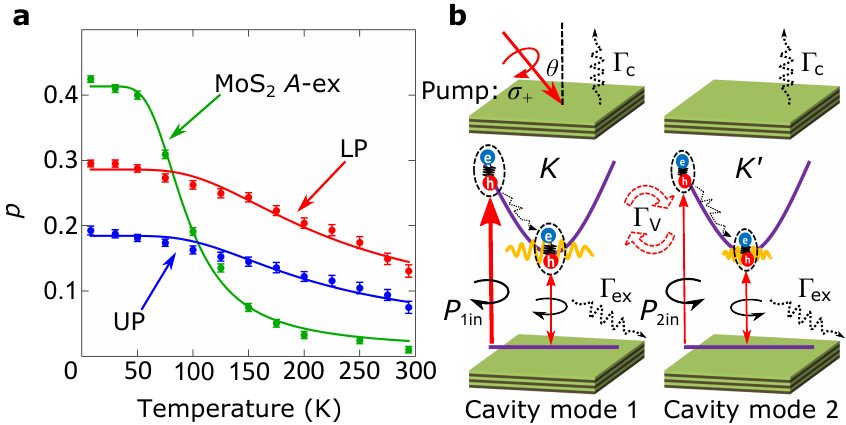}
\caption{\textbf{Temperature-dependent valley polarized exciton-polariton emission.} \textbf{a}, Emission polarization for bare and MC-MoS$_2$ pumped with $E_\textrm{{ph}}$ = 1.938 eV. Lines are fits to the emission rate equation and the valley-sensitive cavity QED model, respectively. \textbf{b}, Schematics of the pumping, intervalley-scattering, exciton-photon coupling, and decay mechanisms present for valley polarized exciton-polaritons in MC-MoS$_2$ with two orthogonal circularly-polarized cavity modes.} \label{figure4}
\end{figure}

Because emission polarization from valley states is determined by the relative exciton relaxation and intervalley scattering rates~\cite{mak2012control,kioseoglou2012valley}, the evolution of $p$ with temperature probes valley scattering dynamics. Fig.~\ref{figure4}a shows the temperature dependence $p(T)$ from both bare MoS$_2$ and MC-MoS$_2$. In bare MoS$_2$, the low-temperature valley polarization $p$ of $A$-exciton vanishes quickly as the temperature is raised above 100~K. This behavior originates from the ratio of the exciton and valley relaxation rates, as predicted by a rate equation model (see Supplementary Information)~\cite{mak2012control,kioseoglou2012valley}:
\begin{equation}
p_{\text{bare}} =  \frac{A_{\rm bare}}{1+\frac{2\Gamma_{\rm v}}{\Gamma_{\rm ex}}}
\label{eq:pol2}
\end{equation}
where $A_{\rm bare}$ is a constant depending on optical pumping, $\Gamma_{\rm ex}$ is the exciton relaxation rate, and $\Gamma_{\rm v}$ is the intervalley scattering rate. The temperature evolution of $p$ can therefore be interpreted as a measurement of the ratio $\Gamma_{\rm v}/\Gamma_{\rm ex}$.  At elevated temperatures, $\Gamma_{\rm v}$ increases due to thermally-activated phonon-assisted intervalley scattering~\cite{kioseoglou2012valley,zeng2012valley} and emission polarization is suppressed.

The temperature-dependent emission polarization for exciton-polaritons in MC-MoS$_2$ is very different from bare MoS$_2$. The low-temperature UP and LP polarizations $p_{\rm UP}$ and $p_{\rm LP}$ are slightly smaller than that in bare MoS$_2$, which is expected due to the non-perfect circular polarization of the cavity field from oblique incidence~(see Supplementary Information), the differential shift of the exciton energy with temperature, and the inhomogeneity of the MoS$_2$ flakes in the collection region of the cavity. At room temperature, the polarizations $p_{\rm UP}(T)$ and $p_{\rm LP}(T)$ decrease smoothly to 7.5$\%$ and 13$\%$, in contrast to the vanishing polarization in bare MoS$_2$.

We explain this difference using a valley-sensitive cavity rate model accounting for coherent exchange and incoherent scattering (Fig.~\ref{figure4}b).  Our phenomenological model is a generalization of the standard steady state master equation approach for interacting boson modes in cavity QED~\cite{laussy2009luminescence}, modified to include two distinct MC photon modes (of different polarization) coupled to two distinct exciton modes (for $K$ and $K^\prime$ valleys). The cavity QED rate equations are augmented with incoherent intervalley scattering between each bosonic valley mode with rate $\Gamma_{\rm v}$ proportional to the exciton population of the other valley, and the exciton and photon modes relax with rates $\Gamma_{\rm ex}$ and $\Gamma_{\rm c}$, respectively. Details of the valley-sensitive cavity QED model are in the Supplementary Information. In the high cooperativity regime ($4g^2 \gg \Gamma_{\rm c}\Gamma_{\rm ex}$), which is necessarily satisfied for the strong coupling regime for the MoS$_2$ MCs, the polarization of the steady state cavity emission takes a convenient form:

\begin{equation}
p_{\text{MC}} = \frac{A_{\rm MC}}{1+\frac{2\Gamma_{\rm v}}{\Gamma_{\rm ex} +\Gamma_{\rm c}}}
\label{eq:pol3}
\end{equation}
Eq.~\eqref{eq:pol3} has the same structure as Eq.~\eqref{eq:pol2} describing a bare monolayer, but relaxation now occurs equally through both exciton and cavity channels. Importantly, only the excitonic part of the exciton-polariton quasiparticle experiences the intervalley scattering while the polarized photonic amplitudes do not couple for a good cavity with negligible mode mixing. The ratio in the denominator of Eq.~\eqref{eq:pol3} is therefore reduced compared to the bare MoS$_2$. Radiative decay of the exciton-polaritons through cavity emission occurs before incoherent valley scattering, and the polarization of the output field persists even at elevated temperatures. Additional details of this model are in the Supplementary Information.

This picture is confirmed from the temperature dependence $p(T)$.
Although the homogeneous exciton radiative lifetime is highly temperature dependent~\cite{Moody2015, Robert2016}, the total exciton decay time in MoS$_2$ has been found to be nearly temperature independent $\tau_{\rm ex} = \hbar/\Gamma_{\rm ex} = 4$~ps~\cite{lagarde2014carrier, Palummo2015}. The intervalley scattering rate is proportional to the phonon thermal population $\Gamma_{\rm v}  \sim \exp\left(-E/k_{\rm B}T\right)$~\cite{zeng2012valley}. These assumptions allow $p_{\rm bare}$ to be well-fit by Eq.~\eqref{eq:pol2} with $A_{\rm bare} = 0.413 \pm 0.009$ and $E = 35 \pm 4$ meV.

 Using this same phonon model parameters as inputs to Eq.~\eqref{eq:pol2} yields a two-parameter fit for $p_{\rm MC}$ with the amplitude $A_{\rm MC}$ and $\Gamma_{\rm c}$ the only free parameters~(Fig.~\ref{figure4}a). The extracted $\Gamma_{\rm c} = 4.2$~meV (UP) and 5.2~meV (LP) are consistent with the intrinsic linewidth (4 meV without angle broadening) of our MC, demonstrating that this valley-sensitive cavity model is a self-consistent description of the exciton-polariton polarization.  The good agreement of the valley-sensitive cavity model with the emission polarization verifies the observation of valley polarized exciton-polaritons with modified dynamics distinct from typical valley excitations in monolayer semiconductors.

The preservation of the polarized emission from light-matter quasiparticles in MoS$_2$ implies that the two classes of exciton-polariton with distinct valley index presented in Fig.~\ref{figure1} can be selectively excited. It is therefore necessary to treat the TMD exciton-polariton system as a having two classes of exciton-polariton, each with upper and lower branches. The optical coherence possible between these distinct, spatially co-existing polaritonic quasiparticles is determined by both the dynamics of intervalley scattering and the photonic cavity.


In summary, we have established the existence of strongly-coupled exciton-polaritons with polarized valley pseudospin in a 2D TMD semiconductor. These hybrid light-matter quasiparticles can be selectively pumped and probed in photonic microcavities using circularly polarized light. The dynamics of these polarization-sensitive quasiparticles are distinct from valley excitons in TMD monolayers due to the cavity photon, representing `half' of the hybrid quasiparticle, being insensitive to intervalley scattering, resulting in preservation of appreciable exciton-polariton valley polarization at room temperature. The capability to select distinct valley-polarized light-matter quasiparticles in cavity QED with circular polarization makes possible a platform to bring the valley and spin degrees of freedom to two-dimensional semiconductor polaritonics~\cite{Low2016, Basov2016}.

\section*{Methods\vspace{-.25em}}

\textbf{Sample preparation.}  The DBR is fabricated by depositing alternating layers of SiO$_2$ and Si$_3$N$_4$ on a silicon wafer substrate using plasma-enhanced chemical vapor deposition (PECVD).  The top DBR has fewer pairs than the bottom to favor emission in the detector direction. The thickness of the inner cavity region of SiO$_2$ between the DBRs corresponds to $\lambda/2n$, with $\lambda$ near the MoS$_2$ $A$-exciton transition wavelength and $n \approx 1.45$ the index of refraction of SiO$_2$.  Monolayer MoS$_2$ is grown onto a SiO$_2$/Si wafer by chemical vapor deposition (CVD) at atmospheric pressure~\cite{lee2012synthesis}. The physical and optical properties of the MoS$_2$ monolayers are characterized through atomic force microscopy, optical microscopy, Raman spectroscopy, and PL measurements (see Supplementary Information). The CVD-grown monolayer was transferred using a polycarbonate film~\cite{lin2011clean,park2010growth} onto the bottom DBR, and the system was then encapsulated by growing a top DBR using PECVD.  Both as-grown and polycarbonate-transferred monolayer MoS$_2$ have a direct-gap transition of the $A$-exciton at 1.855~eV with linewidth around 64~meV. Details of CVD procedures, sample structure, and characterization can be found in the supplementary information. \\

\textbf{Optical measurement.} For the MC-MoS$_2$, angle-resolved reflectivity measurements were  performed using a goniometer with angular resolution of $\ang{1}$. A stabilized Tungsten-Halogen light source (Thorlabs) was used as the white light source. The spot size on the sample is estimated at 200~$\mu$m. The reflected spectrum was analyzed with a fiber-coupled spectrometer and CCD (Andor). For polarized PL measurements, emission was collected with a lens along the normal direction using the same spectrometer. A tunable CW dye laser (Matisse 2DR) with linewidth less than 20~MHz was used to pump the sample from 600 to 645~nm with intensity 9.5~W/cm$^2$ over a spot size of $\sim 200$~$\mu$m. The samples were loaded inside an optical cryostat (Advanced Research Systems) capable of reaching 6~K. For all the data shown, the MC is pumped with $\sigma _{+}$ circularly polarized light and the $\sigma _{+}$ and $\sigma _{-}$ polarizations are resolved in collection. The same conclusions are achieved by pumping with $\sigma _{-}$ light. Reported best-fit parameters are obtained from weighted least-squares fitting with parameter uncertainties estimated using resampling. For polarized PL measurements on the bare MoS$_2$, pumping and collection were performed using an objective (spot size $\sim1 \mu$m) to minimize the inhomogeneous broadening over the CVD monolayer to obtain the intrinsic polarization of the material.\\

\section*{Acknowledgments\vspace{-.25em}}

This research is supported by the U.S. Department of Energy, Office of Basic Energy Sciences, Division of Materials Sciences and Engineering under Award No. DE-SC0012130 (cavity spectroscopy), by the National Science Foundation MRSEC program under Grant No. DMR-1121262 (transfer and device assembly), and by the National Science Foundation under Grant No. DMR-1507810 (CVD growth). This work made use of the EPIC and KECK-II facilities of the NU\emph{ANCE} Center at Northwestern University, which has received support from the Soft and Hybrid Nanotechnology Experimental (SHyNE) Resource (NSF NNCI-1542205); the MRSEC program (NSF DMR-1121262) at the Materials Research Center; the International Institute for Nanotechnology (IIN); the Keck Foundation; and the State of Illinois, through the IIN. This work utilized Northwestern University Micro/Nano Fabrication Facility (NUFAB), which is partially supported by Soft and Hybrid Nanotechnology Experimental (SHyNE) Resource (NSF NNCI-1542205), the Materials Research Science and Engineering Center (NSF DMR-1121262), the State of Illinois, and Northwestern University. J.D.C. is supported by the Department of Defense through the National Defense Science and Engineering Fellowship (NDSEG) Program. J.D.C also gratefully acknowledges support from the Ryan Fellowship and the IIN. N.P.S. acknowledges support as an Alfred P. Sloan Research Fellow.

\section*{Author contributions}
N.P.S. and Y.J.C. conceived the experiments. Y.J.C. fabricated the MC. J.D.C. and V.P.D. synthesized the monolayers. T.K.S. and Y.J.C. performed material characterization and transfer. Y.J.C. performed measurements and modeling. Y.J.C. and N.P.S co-wrote the paper. All authors discussed the results and commented on the manuscript.

\section*{Competing financial interests}
The authors declare no competing financial interests.

\bibliography{Chen_ValleyPolarized_EP}

\begin{thebibliography}{50}%
\makeatletter
\providecommand \@ifxundefined [1]{%
 \@ifx{#1\undefined}
}%
\providecommand \@ifnum [1]{%
 \ifnum #1\expandafter \@firstoftwo
 \else \expandafter \@secondoftwo
 \fi
}%
\providecommand \@ifx [1]{%
 \ifx #1\expandafter \@firstoftwo
 \else \expandafter \@secondoftwo
 \fi
}%
\providecommand \natexlab [1]{#1}%
\providecommand \enquote  [1]{``#1''}%
\providecommand \bibnamefont  [1]{#1}%
\providecommand \bibfnamefont [1]{#1}%
\providecommand \citenamefont [1]{#1}%
\providecommand \href@noop [0]{\@secondoftwo}%
\providecommand \href [0]{\begingroup \@sanitize@url \@href}%
\providecommand \@href[1]{\@@startlink{#1}\@@href}%
\providecommand \@@href[1]{\endgroup#1\@@endlink}%
\providecommand \@sanitize@url [0]{\catcode `\\12\catcode `\$12\catcode
  `\&12\catcode `\#12\catcode `\^12\catcode `\_12\catcode `\%12\relax}%
\providecommand \@@startlink[1]{}%
\providecommand \@@endlink[0]{}%
\providecommand \url  [0]{\begingroup\@sanitize@url \@url }%
\providecommand \@url [1]{\endgroup\@href {#1}{\urlprefix }}%
\providecommand \urlprefix  [0]{URL }%
\providecommand \Eprint [0]{\href }%
\providecommand \doibase [0]{http://dx.doi.org/}%
\providecommand \selectlanguage [0]{\@gobble}%
\providecommand \bibinfo  [0]{\@secondoftwo}%
\providecommand \bibfield  [0]{\@secondoftwo}%
\providecommand \translation [1]{[#1]}%
\providecommand \BibitemOpen [0]{}%
\providecommand \bibitemStop [0]{}%
\providecommand \bibitemNoStop [0]{.\EOS\space}%
\providecommand \EOS [0]{\spacefactor3000\relax}%
\providecommand \BibitemShut  [1]{\csname bibitem#1\endcsname}%
\let\auto@bib@innerbib\@empty
\bibitem [{\citenamefont {Purcell}(1946)}]{Purcell1946}%
  \BibitemOpen
  \bibfield  {author} {\bibinfo {author} {\bibfnamefont {E.~M.}\ \bibnamefont
  {Purcell}},\ }\bibfield  {title} {\enquote {\bibinfo {title} {Spontaneous
  emission probabilities at radio frequencies},}\ }\href {\doibase
  10.1103/PhysRev.69.674.2} {\bibfield  {journal} {\bibinfo  {journal} {Phys.
  Rev.}\ }\textbf {\bibinfo {volume} {69}},\ \bibinfo {pages} {681} (\bibinfo
  {year} {1946})}\BibitemShut {NoStop}%
\bibitem [{\citenamefont {Haroche}\ and\ \citenamefont
  {Kleppner}(1989)}]{Haroche1989}%
  \BibitemOpen
  \bibfield  {author} {\bibinfo {author} {\bibfnamefont {Serge}\ \bibnamefont
  {Haroche}}\ and\ \bibinfo {author} {\bibfnamefont {Daniel}\ \bibnamefont
  {Kleppner}},\ }\bibfield  {title} {\enquote {\bibinfo {title} {Cavity quantum
  electrodynamics},}\ }\href {\doibase 10.1063/1.881201} {\bibfield  {journal}
  {\bibinfo  {journal} {Physics Today}\ }\textbf {\bibinfo {volume} {42}},\
  \bibinfo {pages} {24--30} (\bibinfo {year} {1989})}\BibitemShut {NoStop}%
\bibitem [{\citenamefont {Kimble}(1998)}]{Kimble1998}%
  \BibitemOpen
  \bibfield  {author} {\bibinfo {author} {\bibfnamefont {H~J}\ \bibnamefont
  {Kimble}},\ }\bibfield  {title} {\enquote {\bibinfo {title} {Strong
  interactions of single atoms and photons in cavity {QED}},}\ }\href
  {http://stacks.iop.org/1402-4896/1998/i=T76/a=019} {\bibfield  {journal}
  {\bibinfo  {journal} {Phys. Scr.}\ }\textbf {\bibinfo {volume} {1998}},\
  \bibinfo {pages} {127} (\bibinfo {year} {1998})}\BibitemShut {NoStop}%
\bibitem [{\citenamefont {Reithmaier}\ \emph {et~al.}(2004)\citenamefont
  {Reithmaier}, \citenamefont {Sek}, \citenamefont {Loffler}, \citenamefont
  {Hofmann}, \citenamefont {Kuhn}, \citenamefont {Reitzenstein}, \citenamefont
  {Keldysh}, \citenamefont {Kulakovskii}, \citenamefont {Reinecke},\ and\
  \citenamefont {Forchel}}]{Reithmaier2004}%
  \BibitemOpen
  \bibfield  {author} {\bibinfo {author} {\bibfnamefont {J.~P.}\ \bibnamefont
  {Reithmaier}}, \bibinfo {author} {\bibfnamefont {G.}~\bibnamefont {Sek}},
  \bibinfo {author} {\bibfnamefont {A.}~\bibnamefont {Loffler}}, \bibinfo
  {author} {\bibfnamefont {C.}~\bibnamefont {Hofmann}}, \bibinfo {author}
  {\bibfnamefont {S.}~\bibnamefont {Kuhn}}, \bibinfo {author} {\bibfnamefont
  {S.}~\bibnamefont {Reitzenstein}}, \bibinfo {author} {\bibfnamefont {L.~V.}\
  \bibnamefont {Keldysh}}, \bibinfo {author} {\bibfnamefont {V.~D.}\
  \bibnamefont {Kulakovskii}}, \bibinfo {author} {\bibfnamefont {T.~L.}\
  \bibnamefont {Reinecke}}, \ and\ \bibinfo {author} {\bibfnamefont
  {A.}~\bibnamefont {Forchel}},\ }\bibfield  {title} {\enquote {\bibinfo
  {title} {Strong coupling in a single quantum dot-semiconductor microcavity
  system},}\ }\href {http://dx.doi.org/10.1038/nature02969} {\bibfield
  {journal} {\bibinfo  {journal} {Nature}\ }\textbf {\bibinfo {volume} {432}},\
  \bibinfo {pages} {197--200} (\bibinfo {year} {2004})}\BibitemShut {NoStop}%
\bibitem [{\citenamefont {Yoshie}\ \emph {et~al.}(2004)\citenamefont {Yoshie},
  \citenamefont {Scherer}, \citenamefont {Hendrickson}, \citenamefont
  {Khitrova}, \citenamefont {Gibbs}, \citenamefont {Rupper}, \citenamefont
  {Ell}, \citenamefont {Shchekin},\ and\ \citenamefont {Deppe}}]{Yoshie2004}%
  \BibitemOpen
  \bibfield  {author} {\bibinfo {author} {\bibfnamefont {T.}~\bibnamefont
  {Yoshie}}, \bibinfo {author} {\bibfnamefont {A.}~\bibnamefont {Scherer}},
  \bibinfo {author} {\bibfnamefont {J.}~\bibnamefont {Hendrickson}}, \bibinfo
  {author} {\bibfnamefont {G.}~\bibnamefont {Khitrova}}, \bibinfo {author}
  {\bibfnamefont {H.~M.}\ \bibnamefont {Gibbs}}, \bibinfo {author}
  {\bibfnamefont {G.}~\bibnamefont {Rupper}}, \bibinfo {author} {\bibfnamefont
  {C.}~\bibnamefont {Ell}}, \bibinfo {author} {\bibfnamefont {O.~B.}\
  \bibnamefont {Shchekin}}, \ and\ \bibinfo {author} {\bibfnamefont {D.~G.}\
  \bibnamefont {Deppe}},\ }\bibfield  {title} {\enquote {\bibinfo {title}
  {Vacuum {R}abi splitting with a single quantum dot in a photonic crystal
  nanocavity},}\ }\href {http://dx.doi.org/10.1038/nature03119} {\bibfield
  {journal} {\bibinfo  {journal} {Nature}\ }\textbf {\bibinfo {volume} {432}},\
  \bibinfo {pages} {200--203} (\bibinfo {year} {2004})}\BibitemShut {NoStop}%
\bibitem [{\citenamefont {Wallraff}\ \emph {et~al.}(2004)\citenamefont
  {Wallraff}, \citenamefont {Schuster}, \citenamefont {Blais}, \citenamefont
  {Frunzio}, \citenamefont {Huang}, \citenamefont {Majer}, \citenamefont
  {Kumar}, \citenamefont {Girvin},\ and\ \citenamefont
  {Schoelkopf}}]{Wallraff2004}%
  \BibitemOpen
  \bibfield  {author} {\bibinfo {author} {\bibfnamefont {A.}~\bibnamefont
  {Wallraff}}, \bibinfo {author} {\bibfnamefont {D.~I.}\ \bibnamefont
  {Schuster}}, \bibinfo {author} {\bibfnamefont {A.}~\bibnamefont {Blais}},
  \bibinfo {author} {\bibfnamefont {L.}~\bibnamefont {Frunzio}}, \bibinfo
  {author} {\bibfnamefont {R.-S.}\ \bibnamefont {Huang}}, \bibinfo {author}
  {\bibfnamefont {J.}~\bibnamefont {Majer}}, \bibinfo {author} {\bibfnamefont
  {S.}~\bibnamefont {Kumar}}, \bibinfo {author} {\bibfnamefont {S.~M.}\
  \bibnamefont {Girvin}}, \ and\ \bibinfo {author} {\bibfnamefont {R.~J.}\
  \bibnamefont {Schoelkopf}},\ }\bibfield  {title} {\enquote {\bibinfo {title}
  {Strong coupling of a single photon to a superconducting qubit using circuit
  quantum electrodynamics},}\ }\href {http://dx.doi.org/10.1038/nature02851}
  {\bibfield  {journal} {\bibinfo  {journal} {Nature}\ }\textbf {\bibinfo
  {volume} {431}},\ \bibinfo {pages} {162--167} (\bibinfo {year}
  {2004})}\BibitemShut {NoStop}%
\bibitem [{\citenamefont {Weisbuch}\ \emph {et~al.}(1992)\citenamefont
  {Weisbuch}, \citenamefont {Nishioka}, \citenamefont {Ishikawa},\ and\
  \citenamefont {Arakawa}}]{Weisbuch1992}%
  \BibitemOpen
  \bibfield  {author} {\bibinfo {author} {\bibfnamefont {Claude}\ \bibnamefont
  {Weisbuch}}, \bibinfo {author} {\bibfnamefont {M}~\bibnamefont {Nishioka}},
  \bibinfo {author} {\bibfnamefont {A}~\bibnamefont {Ishikawa}}, \ and\
  \bibinfo {author} {\bibfnamefont {Y}~\bibnamefont {Arakawa}},\ }\bibfield
  {title} {\enquote {\bibinfo {title} {Observation of the coupled
  exciton-photon mode splitting in a semiconductor quantum microcavity},}\
  }\href@noop {} {\bibfield  {journal} {\bibinfo  {journal} {Phys. Rev. Lett.}\
  }\textbf {\bibinfo {volume} {69}},\ \bibinfo {pages} {3314} (\bibinfo {year}
  {1992})}\BibitemShut {NoStop}%
\bibitem [{\citenamefont {Houdr\'e}\ \emph {et~al.}(1994)\citenamefont
  {Houdr\'e}, \citenamefont {Weisbuch}, \citenamefont {Stanley}, \citenamefont
  {Oesterle}, \citenamefont {Pellandini},\ and\ \citenamefont
  {Ilegems}}]{Houdre1994}%
  \BibitemOpen
  \bibfield  {author} {\bibinfo {author} {\bibfnamefont {R.}~\bibnamefont
  {Houdr\'e}}, \bibinfo {author} {\bibfnamefont {C.}~\bibnamefont {Weisbuch}},
  \bibinfo {author} {\bibfnamefont {R.~P.}\ \bibnamefont {Stanley}}, \bibinfo
  {author} {\bibfnamefont {U.}~\bibnamefont {Oesterle}}, \bibinfo {author}
  {\bibfnamefont {P.}~\bibnamefont {Pellandini}}, \ and\ \bibinfo {author}
  {\bibfnamefont {M.}~\bibnamefont {Ilegems}},\ }\bibfield  {title} {\enquote
  {\bibinfo {title} {Measurement of cavity-polariton dispersion curve from
  angle-resolved photoluminescence experiments},}\ }\href {\doibase
  10.1103/PhysRevLett.73.2043} {\bibfield  {journal} {\bibinfo  {journal}
  {Phys. Rev. Lett.}\ }\textbf {\bibinfo {volume} {73}},\ \bibinfo {pages}
  {2043--2046} (\bibinfo {year} {1994})}\BibitemShut {NoStop}%
\bibitem [{\citenamefont {Deng}\ \emph {et~al.}(2010)\citenamefont {Deng},
  \citenamefont {Haug},\ and\ \citenamefont {Yamamoto}}]{deng2010exciton}%
  \BibitemOpen
  \bibfield  {author} {\bibinfo {author} {\bibfnamefont {Hui}\ \bibnamefont
  {Deng}}, \bibinfo {author} {\bibfnamefont {Hartmut}\ \bibnamefont {Haug}}, \
  and\ \bibinfo {author} {\bibfnamefont {Yoshihisa}\ \bibnamefont {Yamamoto}},\
  }\bibfield  {title} {\enquote {\bibinfo {title} {Exciton-polariton
  {B}ose-{E}instein condensation},}\ }\href@noop {} {\bibfield  {journal}
  {\bibinfo  {journal} {Rev. Mod. Phys.}\ }\textbf {\bibinfo {volume} {82}},\
  \bibinfo {pages} {1489} (\bibinfo {year} {2010})}\BibitemShut {NoStop}%
\bibitem [{\citenamefont {Amo}\ \emph {et~al.}(2010)\citenamefont {Amo},
  \citenamefont {Liew}, \citenamefont {Adrados}, \citenamefont {Houdr{\'e}},
  \citenamefont {Giacobino}, \citenamefont {Kavokin},\ and\ \citenamefont
  {Bramati}}]{amo2010exciton}%
  \BibitemOpen
  \bibfield  {author} {\bibinfo {author} {\bibfnamefont {A}~\bibnamefont
  {Amo}}, \bibinfo {author} {\bibfnamefont {TCH}\ \bibnamefont {Liew}},
  \bibinfo {author} {\bibfnamefont {C}~\bibnamefont {Adrados}}, \bibinfo
  {author} {\bibfnamefont {R}~\bibnamefont {Houdr{\'e}}}, \bibinfo {author}
  {\bibfnamefont {E}~\bibnamefont {Giacobino}}, \bibinfo {author}
  {\bibfnamefont {AV}~\bibnamefont {Kavokin}}, \ and\ \bibinfo {author}
  {\bibfnamefont {A}~\bibnamefont {Bramati}},\ }\bibfield  {title} {\enquote
  {\bibinfo {title} {Exciton--polariton spin switches},}\ }\href@noop {}
  {\bibfield  {journal} {\bibinfo  {journal} {Nature Photon.}\ }\textbf
  {\bibinfo {volume} {4}},\ \bibinfo {pages} {361--366} (\bibinfo {year}
  {2010})}\BibitemShut {NoStop}%
\bibitem [{\citenamefont {Sanvitto}\ and\ \citenamefont
  {K\'{e}na-Cohen}(2016)}]{Sanvitto2016}%
  \BibitemOpen
  \bibfield  {author} {\bibinfo {author} {\bibfnamefont {Daniele}\ \bibnamefont
  {Sanvitto}}\ and\ \bibinfo {author} {\bibfnamefont {St\'{e}phane}\
  \bibnamefont {K\'{e}na-Cohen}},\ }\bibfield  {title} {\enquote {\bibinfo
  {title} {The road towards polaritonic devices},}\ }\href
  {http://dx.doi.org/10.1038/nmat4668} {\bibfield  {journal} {\bibinfo
  {journal} {Nature Mater.}\ }\textbf {\bibinfo {volume} {15}},\ \bibinfo
  {pages} {1061--1073} (\bibinfo {year} {2016})}\BibitemShut {NoStop}%
\bibitem [{\citenamefont {Lenferink}\ \emph {et~al.}(2014)\citenamefont
  {Lenferink}, \citenamefont {Wei},\ and\ \citenamefont
  {Stern}}]{Lenferink2014}%
  \BibitemOpen
  \bibfield  {author} {\bibinfo {author} {\bibfnamefont {Erik~J.}\ \bibnamefont
  {Lenferink}}, \bibinfo {author} {\bibfnamefont {Guohua}\ \bibnamefont {Wei}},
  \ and\ \bibinfo {author} {\bibfnamefont {Nathaniel~P.}\ \bibnamefont
  {Stern}},\ }\bibfield  {title} {\enquote {\bibinfo {title} {Coherent optical
  non-reciprocity in axisymmetric resonators},}\ }\href {\doibase
  10.1364/OE.22.016099} {\bibfield  {journal} {\bibinfo  {journal} {Opt.
  Express}\ }\textbf {\bibinfo {volume} {22}},\ \bibinfo {pages} {16099--16111}
  (\bibinfo {year} {2014})}\BibitemShut {NoStop}%
\bibitem [{\citenamefont {Sayrin}\ \emph {et~al.}(2015)\citenamefont {Sayrin},
  \citenamefont {Junge}, \citenamefont {Mitsch}, \citenamefont {Albrecht},
  \citenamefont {O'Shea}, \citenamefont {Schneeweiss}, \citenamefont {Volz},\
  and\ \citenamefont {Rauschenbeutel}}]{Sayrin2015}%
  \BibitemOpen
  \bibfield  {author} {\bibinfo {author} {\bibfnamefont {Cl\'ement}\
  \bibnamefont {Sayrin}}, \bibinfo {author} {\bibfnamefont {Christian}\
  \bibnamefont {Junge}}, \bibinfo {author} {\bibfnamefont {Rudolf}\
  \bibnamefont {Mitsch}}, \bibinfo {author} {\bibfnamefont {Bernhard}\
  \bibnamefont {Albrecht}}, \bibinfo {author} {\bibfnamefont {Danny}\
  \bibnamefont {O'Shea}}, \bibinfo {author} {\bibfnamefont {Philipp}\
  \bibnamefont {Schneeweiss}}, \bibinfo {author} {\bibfnamefont {J\"urgen}\
  \bibnamefont {Volz}}, \ and\ \bibinfo {author} {\bibfnamefont {Arno}\
  \bibnamefont {Rauschenbeutel}},\ }\bibfield  {title} {\enquote {\bibinfo
  {title} {Nanophotonic optical isolator controlled by the internal state of
  cold atoms},}\ }\href {\doibase 10.1103/PhysRevX.5.041036} {\bibfield
  {journal} {\bibinfo  {journal} {Phys. Rev. X}\ }\textbf {\bibinfo {volume}
  {5}},\ \bibinfo {pages} {041036} (\bibinfo {year} {2015})}\BibitemShut
  {NoStop}%
\bibitem [{\citenamefont {S\"{o}llner}\ \emph {et~al.}(2015)\citenamefont
  {S\"{o}llner}, \citenamefont {Mahmoodian}, \citenamefont {Hansen},
  \citenamefont {Midolo}, \citenamefont {Javadi}, \citenamefont
  {Kir\v{s}ansk\.e}, \citenamefont {Pregnolato}, \citenamefont {El-Ella},
  \citenamefont {Lee}, \citenamefont {Song}, \citenamefont {Stobbe},\ and\
  \citenamefont {Lodahl}}]{Sollner2015}%
  \BibitemOpen
  \bibfield  {author} {\bibinfo {author} {\bibfnamefont {Immo}\ \bibnamefont
  {S\"{o}llner}}, \bibinfo {author} {\bibfnamefont {Sahand}\ \bibnamefont
  {Mahmoodian}}, \bibinfo {author} {\bibfnamefont {Sofie~Lindskov}\
  \bibnamefont {Hansen}}, \bibinfo {author} {\bibfnamefont {Leonardo}\
  \bibnamefont {Midolo}}, \bibinfo {author} {\bibfnamefont {Alisa}\
  \bibnamefont {Javadi}}, \bibinfo {author} {\bibfnamefont {Gabija}\
  \bibnamefont {Kir\v{s}ansk\.e}}, \bibinfo {author} {\bibfnamefont {Tommaso}\
  \bibnamefont {Pregnolato}}, \bibinfo {author} {\bibfnamefont {Haitham}\
  \bibnamefont {El-Ella}}, \bibinfo {author} {\bibfnamefont {Eun~Hye}\
  \bibnamefont {Lee}}, \bibinfo {author} {\bibfnamefont {Jin~Dong}\
  \bibnamefont {Song}}, \bibinfo {author} {\bibfnamefont {S{\o}ren}\
  \bibnamefont {Stobbe}}, \ and\ \bibinfo {author} {\bibfnamefont {Peter}\
  \bibnamefont {Lodahl}},\ }\bibfield  {title} {\enquote {\bibinfo {title}
  {Deterministic photon-emitter coupling in chiral photonic circuits},}\ }\href
  {http://dx.doi.org/10.1038/nnano.2015.159} {\bibfield  {journal} {\bibinfo
  {journal} {Nat Nano}\ }\textbf {\bibinfo {volume} {10}},\ \bibinfo {pages}
  {775--778} (\bibinfo {year} {2015})}\BibitemShut {NoStop}%
\bibitem [{\citenamefont {Junge}\ \emph {et~al.}(2013)\citenamefont {Junge},
  \citenamefont {O'Shea}, \citenamefont {Volz},\ and\ \citenamefont
  {Rauschenbeutel}}]{Junge2013}%
  \BibitemOpen
  \bibfield  {author} {\bibinfo {author} {\bibfnamefont {Christian}\
  \bibnamefont {Junge}}, \bibinfo {author} {\bibfnamefont {Danny}\ \bibnamefont
  {O'Shea}}, \bibinfo {author} {\bibfnamefont {J\"urgen}\ \bibnamefont {Volz}},
  \ and\ \bibinfo {author} {\bibfnamefont {Arno}\ \bibnamefont
  {Rauschenbeutel}},\ }\bibfield  {title} {\enquote {\bibinfo {title} {Strong
  coupling between single atoms and nontransversal photons},}\ }\href {\doibase
  10.1103/PhysRevLett.110.213604} {\bibfield  {journal} {\bibinfo  {journal}
  {Phys. Rev. Lett.}\ }\textbf {\bibinfo {volume} {110}},\ \bibinfo {pages}
  {213604} (\bibinfo {year} {2013})}\BibitemShut {NoStop}%
\bibitem [{\citenamefont {Mak}\ \emph {et~al.}(2010)\citenamefont {Mak},
  \citenamefont {Lee}, \citenamefont {Hone}, \citenamefont {Shan},\ and\
  \citenamefont {Heinz}}]{mak2010atomically}%
  \BibitemOpen
  \bibfield  {author} {\bibinfo {author} {\bibfnamefont {Kin~Fai}\ \bibnamefont
  {Mak}}, \bibinfo {author} {\bibfnamefont {Changgu}\ \bibnamefont {Lee}},
  \bibinfo {author} {\bibfnamefont {James}\ \bibnamefont {Hone}}, \bibinfo
  {author} {\bibfnamefont {Jie}\ \bibnamefont {Shan}}, \ and\ \bibinfo {author}
  {\bibfnamefont {Tony~F}\ \bibnamefont {Heinz}},\ }\bibfield  {title}
  {\enquote {\bibinfo {title} {Atomically thin {M}o{S}$_2$: a new direct-gap
  semiconductor},}\ }\href@noop {} {\bibfield  {journal} {\bibinfo  {journal}
  {Phys. Rev. Lett.}\ }\textbf {\bibinfo {volume} {105}},\ \bibinfo {pages}
  {136805} (\bibinfo {year} {2010})}\BibitemShut {NoStop}%
\bibitem [{\citenamefont {Splendiani}\ \emph {et~al.}(2010)\citenamefont
  {Splendiani}, \citenamefont {Sun}, \citenamefont {Zhang}, \citenamefont {Li},
  \citenamefont {Kim}, \citenamefont {Chim}, \citenamefont {Galli},\ and\
  \citenamefont {Wang}}]{splendiani2010emerging}%
  \BibitemOpen
  \bibfield  {author} {\bibinfo {author} {\bibfnamefont {Andrea}\ \bibnamefont
  {Splendiani}}, \bibinfo {author} {\bibfnamefont {Liang}\ \bibnamefont {Sun}},
  \bibinfo {author} {\bibfnamefont {Yuanbo}\ \bibnamefont {Zhang}}, \bibinfo
  {author} {\bibfnamefont {Tianshu}\ \bibnamefont {Li}}, \bibinfo {author}
  {\bibfnamefont {Jonghwan}\ \bibnamefont {Kim}}, \bibinfo {author}
  {\bibfnamefont {Chi-Yung}\ \bibnamefont {Chim}}, \bibinfo {author}
  {\bibfnamefont {Giulia}\ \bibnamefont {Galli}}, \ and\ \bibinfo {author}
  {\bibfnamefont {Feng}\ \bibnamefont {Wang}},\ }\bibfield  {title} {\enquote
  {\bibinfo {title} {Emerging photoluminescence in monolayer {M}o{S}$_2$},}\
  }\href@noop {} {\bibfield  {journal} {\bibinfo  {journal} {Nano Lett.}\
  }\textbf {\bibinfo {volume} {10}},\ \bibinfo {pages} {1271--1275} (\bibinfo
  {year} {2010})}\BibitemShut {NoStop}%
\bibitem [{\citenamefont {Xiao}\ \emph {et~al.}(2012)\citenamefont {Xiao},
  \citenamefont {Liu}, \citenamefont {Feng}, \citenamefont {Xu},\ and\
  \citenamefont {Yao}}]{xiao2012coupled}%
  \BibitemOpen
  \bibfield  {author} {\bibinfo {author} {\bibfnamefont {Di}~\bibnamefont
  {Xiao}}, \bibinfo {author} {\bibfnamefont {Gui-Bin}\ \bibnamefont {Liu}},
  \bibinfo {author} {\bibfnamefont {Wanxiang}\ \bibnamefont {Feng}}, \bibinfo
  {author} {\bibfnamefont {Xiaodong}\ \bibnamefont {Xu}}, \ and\ \bibinfo
  {author} {\bibfnamefont {Wang}\ \bibnamefont {Yao}},\ }\bibfield  {title}
  {\enquote {\bibinfo {title} {Coupled spin and valley physics in monolayers of
  {M}o{S}$_{2}$ and other group-vi dichalcogenides},}\ }\href@noop {}
  {\bibfield  {journal} {\bibinfo  {journal} {Phys. Rev. Lett.}\ }\textbf
  {\bibinfo {volume} {108}},\ \bibinfo {pages} {196802} (\bibinfo {year}
  {2012})}\BibitemShut {NoStop}%
\bibitem [{\citenamefont {Mak}\ \emph {et~al.}(2012)\citenamefont {Mak},
  \citenamefont {He}, \citenamefont {Shan},\ and\ \citenamefont
  {Heinz}}]{mak2012control}%
  \BibitemOpen
  \bibfield  {author} {\bibinfo {author} {\bibfnamefont {Kin~Fai}\ \bibnamefont
  {Mak}}, \bibinfo {author} {\bibfnamefont {Keliang}\ \bibnamefont {He}},
  \bibinfo {author} {\bibfnamefont {Jie}\ \bibnamefont {Shan}}, \ and\ \bibinfo
  {author} {\bibfnamefont {Tony~F}\ \bibnamefont {Heinz}},\ }\bibfield  {title}
  {\enquote {\bibinfo {title} {Control of valley polarization in monolayer
  {M}o{S}$_{2}$ by optical helicity},}\ }\href@noop {} {\bibfield  {journal}
  {\bibinfo  {journal} {Nature Nanotech.}\ }\textbf {\bibinfo {volume} {7}},\
  \bibinfo {pages} {494--498} (\bibinfo {year} {2012})}\BibitemShut {NoStop}%
\bibitem [{\citenamefont {Zeng}\ \emph {et~al.}(2012)\citenamefont {Zeng},
  \citenamefont {Dai}, \citenamefont {Yao}, \citenamefont {Xiao},\ and\
  \citenamefont {Cui}}]{zeng2012valley}%
  \BibitemOpen
  \bibfield  {author} {\bibinfo {author} {\bibfnamefont {Hualing}\ \bibnamefont
  {Zeng}}, \bibinfo {author} {\bibfnamefont {Junfeng}\ \bibnamefont {Dai}},
  \bibinfo {author} {\bibfnamefont {Wang}\ \bibnamefont {Yao}}, \bibinfo
  {author} {\bibfnamefont {Di}~\bibnamefont {Xiao}}, \ and\ \bibinfo {author}
  {\bibfnamefont {Xiaodong}\ \bibnamefont {Cui}},\ }\bibfield  {title}
  {\enquote {\bibinfo {title} {Valley polarization in {M}o{S}$_{2}$ monolayers
  by optical pumping},}\ }\href@noop {} {\bibfield  {journal} {\bibinfo
  {journal} {Nature Nanotech.}\ }\textbf {\bibinfo {volume} {7}},\ \bibinfo
  {pages} {490--493} (\bibinfo {year} {2012})}\BibitemShut {NoStop}%
\bibitem [{\citenamefont {Cao}\ \emph {et~al.}(2012)\citenamefont {Cao},
  \citenamefont {Wang}, \citenamefont {Han}, \citenamefont {Ye}, \citenamefont
  {Zhu}, \citenamefont {Shi}, \citenamefont {Niu}, \citenamefont {Tan},
  \citenamefont {Wang}, \citenamefont {Liu},\ and\ \citenamefont
  {Feng}}]{cao2012valley}%
  \BibitemOpen
  \bibfield  {author} {\bibinfo {author} {\bibfnamefont {Ting}\ \bibnamefont
  {Cao}}, \bibinfo {author} {\bibfnamefont {Gang}\ \bibnamefont {Wang}},
  \bibinfo {author} {\bibfnamefont {Wenpeng}\ \bibnamefont {Han}}, \bibinfo
  {author} {\bibfnamefont {Huiqi}\ \bibnamefont {Ye}}, \bibinfo {author}
  {\bibfnamefont {Chuanrui}\ \bibnamefont {Zhu}}, \bibinfo {author}
  {\bibfnamefont {Junren}\ \bibnamefont {Shi}}, \bibinfo {author}
  {\bibfnamefont {Qian}\ \bibnamefont {Niu}}, \bibinfo {author} {\bibfnamefont
  {Pingheng}\ \bibnamefont {Tan}}, \bibinfo {author} {\bibfnamefont {Enge}\
  \bibnamefont {Wang}}, \bibinfo {author} {\bibfnamefont {Baoli}\ \bibnamefont
  {Liu}}, \ and\ \bibinfo {author} {\bibfnamefont {Ji}~\bibnamefont {Feng}},\
  }\bibfield  {title} {\enquote {\bibinfo {title} {Valley-selective circular
  dichroism of monolayer molybdenum disulphide},}\ }\href@noop {} {\bibfield
  {journal} {\bibinfo  {journal} {Nature Commun.}\ }\textbf {\bibinfo {volume}
  {3}},\ \bibinfo {pages} {887} (\bibinfo {year} {2012})}\BibitemShut {NoStop}%
\bibitem [{\citenamefont {Kioseoglou}\ \emph {et~al.}(2012)\citenamefont
  {Kioseoglou}, \citenamefont {Hanbicki}, \citenamefont {Currie}, \citenamefont
  {Friedman}, \citenamefont {Gunlycke},\ and\ \citenamefont
  {Jonker}}]{kioseoglou2012valley}%
  \BibitemOpen
  \bibfield  {author} {\bibinfo {author} {\bibfnamefont {G}~\bibnamefont
  {Kioseoglou}}, \bibinfo {author} {\bibfnamefont {AT}~\bibnamefont
  {Hanbicki}}, \bibinfo {author} {\bibfnamefont {M}~\bibnamefont {Currie}},
  \bibinfo {author} {\bibfnamefont {AL}~\bibnamefont {Friedman}}, \bibinfo
  {author} {\bibfnamefont {D}~\bibnamefont {Gunlycke}}, \ and\ \bibinfo
  {author} {\bibfnamefont {BT}~\bibnamefont {Jonker}},\ }\bibfield  {title}
  {\enquote {\bibinfo {title} {Valley polarization and intervalley scattering
  in monolayer {M}o{S}$_2$},}\ }\href@noop {} {\bibfield  {journal} {\bibinfo
  {journal} {Appl. Phys. Lett.}\ }\textbf {\bibinfo {volume} {101}},\ \bibinfo
  {pages} {221907} (\bibinfo {year} {2012})}\BibitemShut {NoStop}%
\bibitem [{\citenamefont {Wang}\ \emph {et~al.}(2012)\citenamefont {Wang},
  \citenamefont {Kalantar-Zadeh}, \citenamefont {Kis}, \citenamefont
  {Coleman},\ and\ \citenamefont {Strano}}]{wang2012electronics}%
  \BibitemOpen
  \bibfield  {author} {\bibinfo {author} {\bibfnamefont {Qing~Hua}\
  \bibnamefont {Wang}}, \bibinfo {author} {\bibfnamefont {Kourosh}\
  \bibnamefont {Kalantar-Zadeh}}, \bibinfo {author} {\bibfnamefont {Andras}\
  \bibnamefont {Kis}}, \bibinfo {author} {\bibfnamefont {Jonathan~N}\
  \bibnamefont {Coleman}}, \ and\ \bibinfo {author} {\bibfnamefont {Michael~S}\
  \bibnamefont {Strano}},\ }\bibfield  {title} {\enquote {\bibinfo {title}
  {Electronics and optoelectronics of two-dimensional transition metal
  dichalcogenides},}\ }\href@noop {} {\bibfield  {journal} {\bibinfo  {journal}
  {Nature Nanotech.}\ }\textbf {\bibinfo {volume} {7}},\ \bibinfo {pages}
  {699--712} (\bibinfo {year} {2012})}\BibitemShut {NoStop}%
\bibitem [{\citenamefont {Sallen}\ \emph {et~al.}(2012)\citenamefont {Sallen},
  \citenamefont {Bouet}, \citenamefont {Marie}, \citenamefont {Wang},
  \citenamefont {Zhu}, \citenamefont {Han}, \citenamefont {Lu}, \citenamefont
  {Tan}, \citenamefont {Amand}, \citenamefont {Liu},\ and\ \citenamefont
  {Urbaszek}}]{sallen2012robust}%
  \BibitemOpen
  \bibfield  {author} {\bibinfo {author} {\bibfnamefont {G}~\bibnamefont
  {Sallen}}, \bibinfo {author} {\bibfnamefont {L}~\bibnamefont {Bouet}},
  \bibinfo {author} {\bibfnamefont {X}~\bibnamefont {Marie}}, \bibinfo {author}
  {\bibfnamefont {G}~\bibnamefont {Wang}}, \bibinfo {author} {\bibfnamefont
  {CR}~\bibnamefont {Zhu}}, \bibinfo {author} {\bibfnamefont {WP}~\bibnamefont
  {Han}}, \bibinfo {author} {\bibfnamefont {Y}~\bibnamefont {Lu}}, \bibinfo
  {author} {\bibfnamefont {PH}~\bibnamefont {Tan}}, \bibinfo {author}
  {\bibfnamefont {T}~\bibnamefont {Amand}}, \bibinfo {author} {\bibfnamefont
  {BL}~\bibnamefont {Liu}}, \ and\ \bibinfo {author} {\bibfnamefont
  {B}~\bibnamefont {Urbaszek}},\ }\bibfield  {title} {\enquote {\bibinfo
  {title} {Robust optical emission polarization in mos 2 monolayers through
  selective valley excitation},}\ }\href@noop {} {\bibfield  {journal}
  {\bibinfo  {journal} {Phys. Rev. B}\ }\textbf {\bibinfo {volume} {86}},\
  \bibinfo {pages} {081301} (\bibinfo {year} {2012})}\BibitemShut {NoStop}%
\bibitem [{\citenamefont {Mak}\ \emph {et~al.}(2014)\citenamefont {Mak},
  \citenamefont {McGill}, \citenamefont {Park},\ and\ \citenamefont
  {McEuen}}]{Mak2014}%
  \BibitemOpen
  \bibfield  {author} {\bibinfo {author} {\bibfnamefont {K.~F.}\ \bibnamefont
  {Mak}}, \bibinfo {author} {\bibfnamefont {K.~L.}\ \bibnamefont {McGill}},
  \bibinfo {author} {\bibfnamefont {J.}~\bibnamefont {Park}}, \ and\ \bibinfo
  {author} {\bibfnamefont {P.~L.}\ \bibnamefont {McEuen}},\ }\bibfield  {title}
  {\enquote {\bibinfo {title} {The valley hall effect in {M}o{S}$_2$
  transistors},}\ }\href {\doibase 10.1126/science.1250140} {\bibfield
  {journal} {\bibinfo  {journal} {Science}\ }\textbf {\bibinfo {volume}
  {344}},\ \bibinfo {pages} {1489--1492} (\bibinfo {year} {2014})}\BibitemShut
  {NoStop}%
\bibitem [{\citenamefont {Rohling}\ and\ \citenamefont
  {Burkard}(2012)}]{Rohling2012}%
  \BibitemOpen
  \bibfield  {author} {\bibinfo {author} {\bibfnamefont {Niklas}\ \bibnamefont
  {Rohling}}\ and\ \bibinfo {author} {\bibfnamefont {Guido}\ \bibnamefont
  {Burkard}},\ }\bibfield  {title} {\enquote {\bibinfo {title} {Universal
  quantum computing with spin and valley states},}\ }\href
  {http://stacks.iop.org/1367-2630/14/i=8/a=083008} {\bibfield  {journal}
  {\bibinfo  {journal} {New J. Phys.}\ }\textbf {\bibinfo {volume} {14}},\
  \bibinfo {pages} {083008} (\bibinfo {year} {2012})}\BibitemShut {NoStop}%
\bibitem [{\citenamefont {Behnia}(2012)}]{Behnia2012}%
  \BibitemOpen
  \bibfield  {author} {\bibinfo {author} {\bibfnamefont {Kamran}\ \bibnamefont
  {Behnia}},\ }\bibfield  {title} {\enquote {\bibinfo {title} {Condensed-matter
  physics: Polarized light boosts valleytronics},}\ }\href
  {http://dx.doi.org/10.1038/nnano.2012.117} {\bibfield  {journal} {\bibinfo
  {journal} {Nature Nano.}\ }\textbf {\bibinfo {volume} {7}},\ \bibinfo {pages}
  {488--489} (\bibinfo {year} {2012})}\BibitemShut {NoStop}%
\bibitem [{\citenamefont {Jones}\ \emph {et~al.}(2013)\citenamefont {Jones},
  \citenamefont {Yu}, \citenamefont {Ghimire}, \citenamefont {Wu},
  \citenamefont {Aivazian}, \citenamefont {Ross}, \citenamefont {Zhao},
  \citenamefont {Yan}, \citenamefont {Mandrus}, \citenamefont {Xiao},
  \citenamefont {Yao},\ and\ \citenamefont {Xu}}]{Jones2013}%
  \BibitemOpen
  \bibfield  {author} {\bibinfo {author} {\bibfnamefont {Aaron~M.}\
  \bibnamefont {Jones}}, \bibinfo {author} {\bibfnamefont {Hongyi}\
  \bibnamefont {Yu}}, \bibinfo {author} {\bibfnamefont {Nirmal~J.}\
  \bibnamefont {Ghimire}}, \bibinfo {author} {\bibfnamefont {Sanfeng}\
  \bibnamefont {Wu}}, \bibinfo {author} {\bibfnamefont {Grant}\ \bibnamefont
  {Aivazian}}, \bibinfo {author} {\bibfnamefont {Jason~S.}\ \bibnamefont
  {Ross}}, \bibinfo {author} {\bibfnamefont {Bo}~\bibnamefont {Zhao}}, \bibinfo
  {author} {\bibfnamefont {Jiaqiang}\ \bibnamefont {Yan}}, \bibinfo {author}
  {\bibfnamefont {David~G.}\ \bibnamefont {Mandrus}}, \bibinfo {author}
  {\bibfnamefont {Di}~\bibnamefont {Xiao}}, \bibinfo {author} {\bibfnamefont
  {Wang}\ \bibnamefont {Yao}}, \ and\ \bibinfo {author} {\bibfnamefont
  {Xiaodong}\ \bibnamefont {Xu}},\ }\bibfield  {title} {\enquote {\bibinfo
  {title} {Optical generation of excitonic valley coherence in monolayer
  {WS}e$_2$},}\ }\href {http://dx.doi.org/10.1038/nnano.2013.151} {\bibfield
  {journal} {\bibinfo  {journal} {Nature Nano.}\ }\textbf {\bibinfo {volume}
  {8}},\ \bibinfo {pages} {634--638} (\bibinfo {year} {2013})}\BibitemShut
  {NoStop}%
\bibitem [{\citenamefont {Wang}\ \emph {et~al.}(2016)\citenamefont {Wang},
  \citenamefont {Marie}, \citenamefont {Liu}, \citenamefont {Amand},
  \citenamefont {Robert}, \citenamefont {Cadiz}, \citenamefont {Renucci},\ and\
  \citenamefont {Urbaszek}}]{Wang2016}%
  \BibitemOpen
  \bibfield  {author} {\bibinfo {author} {\bibfnamefont {G.}~\bibnamefont
  {Wang}}, \bibinfo {author} {\bibfnamefont {X.}~\bibnamefont {Marie}},
  \bibinfo {author} {\bibfnamefont {B.~L.}\ \bibnamefont {Liu}}, \bibinfo
  {author} {\bibfnamefont {T.}~\bibnamefont {Amand}}, \bibinfo {author}
  {\bibfnamefont {C.}~\bibnamefont {Robert}}, \bibinfo {author} {\bibfnamefont
  {F.}~\bibnamefont {Cadiz}}, \bibinfo {author} {\bibfnamefont
  {P.}~\bibnamefont {Renucci}}, \ and\ \bibinfo {author} {\bibfnamefont
  {B.}~\bibnamefont {Urbaszek}},\ }\bibfield  {title} {\enquote {\bibinfo
  {title} {Control of exciton valley coherence in transition metal
  dichalcogenide monolayers},}\ }\href {\doibase
  10.1103/PhysRevLett.117.187401} {\bibfield  {journal} {\bibinfo  {journal}
  {Phys. Rev. Lett.}\ }\textbf {\bibinfo {volume} {117}},\ \bibinfo {pages}
  {187401} (\bibinfo {year} {2016})}\BibitemShut {NoStop}%
\bibitem [{\citenamefont {Gan}\ \emph {et~al.}(2013)\citenamefont {Gan},
  \citenamefont {Gao}, \citenamefont {Mak}, \citenamefont {Yao}, \citenamefont
  {Shiue}, \citenamefont {van~der Zande}, \citenamefont {Trusheim},
  \citenamefont {Hatami}, \citenamefont {Heinz}, \citenamefont {Hone},\ and\
  \citenamefont {Englund}}]{gan2013controlling}%
  \BibitemOpen
  \bibfield  {author} {\bibinfo {author} {\bibfnamefont {Xuetao}\ \bibnamefont
  {Gan}}, \bibinfo {author} {\bibfnamefont {Yuanda}\ \bibnamefont {Gao}},
  \bibinfo {author} {\bibfnamefont {Kin~Fai}\ \bibnamefont {Mak}}, \bibinfo
  {author} {\bibfnamefont {Xinwen}\ \bibnamefont {Yao}}, \bibinfo {author}
  {\bibfnamefont {Ren-Jye}\ \bibnamefont {Shiue}}, \bibinfo {author}
  {\bibfnamefont {Arend}\ \bibnamefont {van~der Zande}}, \bibinfo {author}
  {\bibfnamefont {Matthew~E}\ \bibnamefont {Trusheim}}, \bibinfo {author}
  {\bibfnamefont {Fariba}\ \bibnamefont {Hatami}}, \bibinfo {author}
  {\bibfnamefont {Tony~F}\ \bibnamefont {Heinz}}, \bibinfo {author}
  {\bibfnamefont {James}\ \bibnamefont {Hone}}, \ and\ \bibinfo {author}
  {\bibfnamefont {Dirk}\ \bibnamefont {Englund}},\ }\bibfield  {title}
  {\enquote {\bibinfo {title} {Controlling the spontaneous emission rate of
  monolayer {M}o{S}$_2$ in a photonic crystal nanocavity},}\ }\href@noop {}
  {\bibfield  {journal} {\bibinfo  {journal} {Appl. Phys. Lett.}\ }\textbf
  {\bibinfo {volume} {103}},\ \bibinfo {pages} {181119} (\bibinfo {year}
  {2013})}\BibitemShut {NoStop}%
\bibitem [{\citenamefont {Wu}\ \emph {et~al.}(2014)\citenamefont {Wu},
  \citenamefont {Buckley}, \citenamefont {Jones}, \citenamefont {Ross},
  \citenamefont {Ghimire}, \citenamefont {Yan}, \citenamefont {Mandrus},
  \citenamefont {Yao}, \citenamefont {Hatami}, \citenamefont
  {Vu{\v{c}}kovi{\'c}}, \citenamefont {Majumdar},\ and\ \citenamefont
  {Xu}}]{wu2014control}%
  \BibitemOpen
  \bibfield  {author} {\bibinfo {author} {\bibfnamefont {Sanfeng}\ \bibnamefont
  {Wu}}, \bibinfo {author} {\bibfnamefont {Sonia}\ \bibnamefont {Buckley}},
  \bibinfo {author} {\bibfnamefont {Aaron~M}\ \bibnamefont {Jones}}, \bibinfo
  {author} {\bibfnamefont {Jason~S}\ \bibnamefont {Ross}}, \bibinfo {author}
  {\bibfnamefont {Nirmal~J}\ \bibnamefont {Ghimire}}, \bibinfo {author}
  {\bibfnamefont {Jiaqiang}\ \bibnamefont {Yan}}, \bibinfo {author}
  {\bibfnamefont {David~G}\ \bibnamefont {Mandrus}}, \bibinfo {author}
  {\bibfnamefont {Wang}\ \bibnamefont {Yao}}, \bibinfo {author} {\bibfnamefont
  {Fariba}\ \bibnamefont {Hatami}}, \bibinfo {author} {\bibfnamefont {Jelena}\
  \bibnamefont {Vu{\v{c}}kovi{\'c}}}, \bibinfo {author} {\bibfnamefont {Arka}\
  \bibnamefont {Majumdar}}, \ and\ \bibinfo {author} {\bibfnamefont {Xiaodong}\
  \bibnamefont {Xu}},\ }\bibfield  {title} {\enquote {\bibinfo {title} {Control
  of two-dimensional excitonic light emission via photonic crystal},}\
  }\href@noop {} {\bibfield  {journal} {\bibinfo  {journal} {2D Materials}\
  }\textbf {\bibinfo {volume} {1}},\ \bibinfo {pages} {011001} (\bibinfo {year}
  {2014})}\BibitemShut {NoStop}%
\bibitem [{\citenamefont {Schwarz}\ \emph {et~al.}(2014)\citenamefont
  {Schwarz}, \citenamefont {Dufferwiel}, \citenamefont {Walker}, \citenamefont
  {Withers}, \citenamefont {Trichet}, \citenamefont {Sich}, \citenamefont {Li},
  \citenamefont {Chekhovich}, \citenamefont {Borisenko}, \citenamefont
  {Kolesnikov}, \citenamefont {Novoselov}, \citenamefont {Skolnick},
  \citenamefont {Smith}, \citenamefont {Krizhanovskii},\ and\ \citenamefont
  {Tartakovskii}}]{schwarz2014two}%
  \BibitemOpen
  \bibfield  {author} {\bibinfo {author} {\bibfnamefont {Stefan}\ \bibnamefont
  {Schwarz}}, \bibinfo {author} {\bibfnamefont {Scott}\ \bibnamefont
  {Dufferwiel}}, \bibinfo {author} {\bibfnamefont {PM}~\bibnamefont {Walker}},
  \bibinfo {author} {\bibfnamefont {Freddie}\ \bibnamefont {Withers}}, \bibinfo
  {author} {\bibfnamefont {AAP}\ \bibnamefont {Trichet}}, \bibinfo {author}
  {\bibfnamefont {Maksym}\ \bibnamefont {Sich}}, \bibinfo {author}
  {\bibfnamefont {Feng}\ \bibnamefont {Li}}, \bibinfo {author} {\bibfnamefont
  {EA}~\bibnamefont {Chekhovich}}, \bibinfo {author} {\bibfnamefont
  {DN}~\bibnamefont {Borisenko}}, \bibinfo {author} {\bibfnamefont {Nikolai~N}\
  \bibnamefont {Kolesnikov}}, \bibinfo {author} {\bibfnamefont {K.~S.}\
  \bibnamefont {Novoselov}}, \bibinfo {author} {\bibfnamefont {M.~S.}\
  \bibnamefont {Skolnick}}, \bibinfo {author} {\bibfnamefont {J.~M.}\
  \bibnamefont {Smith}}, \bibinfo {author} {\bibfnamefont {D.~N.}\ \bibnamefont
  {Krizhanovskii}}, \ and\ \bibinfo {author} {\bibfnamefont {A.~I.}\
  \bibnamefont {Tartakovskii}},\ }\bibfield  {title} {\enquote {\bibinfo
  {title} {Two-dimensional metal--chalcogenide films in tunable optical
  microcavities},}\ }\href@noop {} {\bibfield  {journal} {\bibinfo  {journal}
  {Nano Lett.}\ }\textbf {\bibinfo {volume} {14}},\ \bibinfo {pages}
  {7003--7008} (\bibinfo {year} {2014})}\BibitemShut {NoStop}%
\bibitem [{\citenamefont {Wei}\ \emph {et~al.}(2015)\citenamefont {Wei},
  \citenamefont {Stanev}, \citenamefont {Czaplewski}, \citenamefont {Jung},\
  and\ \citenamefont {Stern}}]{Wei2015}%
  \BibitemOpen
  \bibfield  {author} {\bibinfo {author} {\bibfnamefont {Guohua}\ \bibnamefont
  {Wei}}, \bibinfo {author} {\bibfnamefont {Teodor~K.}\ \bibnamefont {Stanev}},
  \bibinfo {author} {\bibfnamefont {David~A.}\ \bibnamefont {Czaplewski}},
  \bibinfo {author} {\bibfnamefont {Il~Woong}\ \bibnamefont {Jung}}, \ and\
  \bibinfo {author} {\bibfnamefont {Nathaniel~P.}\ \bibnamefont {Stern}},\
  }\bibfield  {title} {\enquote {\bibinfo {title} {Silicon-nitride photonic
  circuits interfaced with monolayer {MoS$_2$}},}\ }\href {\doibase
  http://dx.doi.org/10.1063/1.4929779} {\bibfield  {journal} {\bibinfo
  {journal} {Appl. Phys. Lett.}\ }\textbf {\bibinfo {volume} {107}},\ \bibinfo
  {eid} {091112} (\bibinfo {year} {2015})}\BibitemShut {NoStop}%
\bibitem [{\citenamefont {Liu}\ \emph {et~al.}(2015)\citenamefont {Liu},
  \citenamefont {Galfsky}, \citenamefont {Sun}, \citenamefont {Xia},
  \citenamefont {Lin}, \citenamefont {Lee}, \citenamefont {K{\'e}na-Cohen},\
  and\ \citenamefont {Menon}}]{liu2015strong}%
  \BibitemOpen
  \bibfield  {author} {\bibinfo {author} {\bibfnamefont {Xiaoze}\ \bibnamefont
  {Liu}}, \bibinfo {author} {\bibfnamefont {Tal}\ \bibnamefont {Galfsky}},
  \bibinfo {author} {\bibfnamefont {Zheng}\ \bibnamefont {Sun}}, \bibinfo
  {author} {\bibfnamefont {Fengnian}\ \bibnamefont {Xia}}, \bibinfo {author}
  {\bibfnamefont {Erh-chen}\ \bibnamefont {Lin}}, \bibinfo {author}
  {\bibfnamefont {Yi-Hsien}\ \bibnamefont {Lee}}, \bibinfo {author}
  {\bibfnamefont {St{\'e}phane}\ \bibnamefont {K{\'e}na-Cohen}}, \ and\
  \bibinfo {author} {\bibfnamefont {Vinod~M}\ \bibnamefont {Menon}},\
  }\bibfield  {title} {\enquote {\bibinfo {title} {Strong light--matter
  coupling in two-dimensional atomic crystals},}\ }\href@noop {} {\bibfield
  {journal} {\bibinfo  {journal} {Nature Photon.}\ }\textbf {\bibinfo {volume}
  {9}},\ \bibinfo {pages} {30--34} (\bibinfo {year} {2015})}\BibitemShut
  {NoStop}%
\bibitem [{\citenamefont {Dufferwiel}\ \emph {et~al.}(2015)\citenamefont
  {Dufferwiel}, \citenamefont {Schwarz}, \citenamefont {Withers}, \citenamefont
  {Trichet}, \citenamefont {Li}, \citenamefont {Sich}, \citenamefont {Del
  Pozo-Zamudio}, \citenamefont {Clark}, \citenamefont {Nalitov}, \citenamefont
  {Solnyshkov}, \citenamefont {Malpuech}, \citenamefont {Novoselov},
  \citenamefont {Smith}, \citenamefont {Skolnick}, \citenamefont
  {Krizhanovskii},\ and\ \citenamefont {Tartakovskii}}]{dufferwiel2015exciton}%
  \BibitemOpen
  \bibfield  {author} {\bibinfo {author} {\bibfnamefont {S}~\bibnamefont
  {Dufferwiel}}, \bibinfo {author} {\bibfnamefont {S}~\bibnamefont {Schwarz}},
  \bibinfo {author} {\bibfnamefont {F}~\bibnamefont {Withers}}, \bibinfo
  {author} {\bibfnamefont {AAP}\ \bibnamefont {Trichet}}, \bibinfo {author}
  {\bibfnamefont {F}~\bibnamefont {Li}}, \bibinfo {author} {\bibfnamefont
  {M}~\bibnamefont {Sich}}, \bibinfo {author} {\bibfnamefont {O}~\bibnamefont
  {Del Pozo-Zamudio}}, \bibinfo {author} {\bibfnamefont {C}~\bibnamefont
  {Clark}}, \bibinfo {author} {\bibfnamefont {A}~\bibnamefont {Nalitov}},
  \bibinfo {author} {\bibfnamefont {DD}~\bibnamefont {Solnyshkov}}, \bibinfo
  {author} {\bibfnamefont {G}~\bibnamefont {Malpuech}}, \bibinfo {author}
  {\bibfnamefont {KS}~\bibnamefont {Novoselov}}, \bibinfo {author}
  {\bibfnamefont {JM}~\bibnamefont {Smith}}, \bibinfo {author} {\bibfnamefont
  {MS}~\bibnamefont {Skolnick}}, \bibinfo {author} {\bibfnamefont
  {DN}~\bibnamefont {Krizhanovskii}}, \ and\ \bibinfo {author} {\bibfnamefont
  {AI}~\bibnamefont {Tartakovskii}},\ }\bibfield  {title} {\enquote {\bibinfo
  {title} {Exciton-polaritons in van der {W}aals heterostructures embedded in
  tunable microcavities},}\ }\href@noop {} {\bibfield  {journal} {\bibinfo
  {journal} {Nat. Commun.}\ }\textbf {\bibinfo {volume} {6}} (\bibinfo {year}
  {2015})}\BibitemShut {NoStop}%
\bibitem [{\citenamefont {Flatten}\ \emph {et~al.}(2016)\citenamefont
  {Flatten}, \citenamefont {He}, \citenamefont {Coles}, \citenamefont
  {Trichet}, \citenamefont {Powell}, \citenamefont {Taylor}, \citenamefont
  {Warner},\ and\ \citenamefont {Smith}}]{Flatten2016}%
  \BibitemOpen
  \bibfield  {author} {\bibinfo {author} {\bibfnamefont {L.~C.}\ \bibnamefont
  {Flatten}}, \bibinfo {author} {\bibfnamefont {Z.}~\bibnamefont {He}},
  \bibinfo {author} {\bibfnamefont {D.~M.}\ \bibnamefont {Coles}}, \bibinfo
  {author} {\bibfnamefont {A.~A.~P.}\ \bibnamefont {Trichet}}, \bibinfo
  {author} {\bibfnamefont {A.~W.}\ \bibnamefont {Powell}}, \bibinfo {author}
  {\bibfnamefont {R.~A.}\ \bibnamefont {Taylor}}, \bibinfo {author}
  {\bibfnamefont {J.~H.}\ \bibnamefont {Warner}}, \ and\ \bibinfo {author}
  {\bibfnamefont {J.~M.}\ \bibnamefont {Smith}},\ }\bibfield  {title} {\enquote
  {\bibinfo {title} {Room-temperature exciton-polaritons with two-dimensional
  ws2},}\ }\href {http://dx.doi.org/10.1038/srep33134} {\bibfield  {journal}
  {\bibinfo  {journal} {Scientific Reports}\ }\textbf {\bibinfo {volume} {6}},\
  \bibinfo {pages} {33134} (\bibinfo {year} {2016})}\BibitemShut {NoStop}%
\bibitem [{\citenamefont {Lundt}\ \emph {et~al.}(2017)\citenamefont {Lundt},
  \citenamefont {Mary\'{n}ski}, \citenamefont {Cherotchenko}, \citenamefont
  {Pant}, \citenamefont {Fan}, \citenamefont {Tongay}, \citenamefont {S\c{e}k},
  \citenamefont {Kavokin}, \citenamefont {H\"{o}fling},\ and\ \citenamefont
  {Schneider}}]{Lundt2017}%
  \BibitemOpen
  \bibfield  {author} {\bibinfo {author} {\bibfnamefont {N}~\bibnamefont
  {Lundt}}, \bibinfo {author} {\bibfnamefont {A}~\bibnamefont {Mary\'{n}ski}},
  \bibinfo {author} {\bibfnamefont {E}~\bibnamefont {Cherotchenko}}, \bibinfo
  {author} {\bibfnamefont {A}~\bibnamefont {Pant}}, \bibinfo {author}
  {\bibfnamefont {X}~\bibnamefont {Fan}}, \bibinfo {author} {\bibfnamefont
  {S}~\bibnamefont {Tongay}}, \bibinfo {author} {\bibfnamefont {G}~\bibnamefont
  {S\c{e}k}}, \bibinfo {author} {\bibfnamefont {A~V}\ \bibnamefont {Kavokin}},
  \bibinfo {author} {\bibfnamefont {S}~\bibnamefont {H\"{o}fling}}, \ and\
  \bibinfo {author} {\bibfnamefont {C}~\bibnamefont {Schneider}},\ }\bibfield
  {title} {\enquote {\bibinfo {title} {Monolayered {M}o{S}e$_2$ : a candidate
  for room temperature polaritonics},}\ }\href
  {http://stacks.iop.org/2053-1583/4/i=1/a=015006} {\bibfield  {journal}
  {\bibinfo  {journal} {2D Materials}\ }\textbf {\bibinfo {volume} {4}},\
  \bibinfo {pages} {015006} (\bibinfo {year} {2017})}\BibitemShut {NoStop}%
\bibitem [{\citenamefont {Low}\ \emph {et~al.}(2016)\citenamefont {Low},
  \citenamefont {Chaves}, \citenamefont {Caldwell}, \citenamefont {Kumar},
  \citenamefont {Fang}, \citenamefont {Avouris}, \citenamefont {Heinz},
  \citenamefont {Guinea}, \citenamefont {Martin-Moreno},\ and\ \citenamefont
  {Koppens}}]{Low2016}%
  \BibitemOpen
  \bibfield  {author} {\bibinfo {author} {\bibfnamefont {Tony}\ \bibnamefont
  {Low}}, \bibinfo {author} {\bibfnamefont {Andrey}\ \bibnamefont {Chaves}},
  \bibinfo {author} {\bibfnamefont {Joshua~D.}\ \bibnamefont {Caldwell}},
  \bibinfo {author} {\bibfnamefont {Anshuman}\ \bibnamefont {Kumar}}, \bibinfo
  {author} {\bibfnamefont {Nicholas~X.}\ \bibnamefont {Fang}}, \bibinfo
  {author} {\bibfnamefont {Phaedon}\ \bibnamefont {Avouris}}, \bibinfo {author}
  {\bibfnamefont {Tony~F.}\ \bibnamefont {Heinz}}, \bibinfo {author}
  {\bibfnamefont {Francisco}\ \bibnamefont {Guinea}}, \bibinfo {author}
  {\bibfnamefont {Luis}\ \bibnamefont {Martin-Moreno}}, \ and\ \bibinfo
  {author} {\bibfnamefont {Frank}\ \bibnamefont {Koppens}},\ }\bibfield
  {title} {\enquote {\bibinfo {title} {Polaritons in layered two-dimensional
  materials},}\ }\href {http://dx.doi.org/10.1038/nmat4792} {\bibfield
  {journal} {\bibinfo  {journal} {Nat Mater}\ }\textbf {\bibinfo {volume}
  {advance online publication}} (\bibinfo {year} {2016})}\BibitemShut {NoStop}%
\bibitem [{\citenamefont {Basov}\ \emph {et~al.}(2016)\citenamefont {Basov},
  \citenamefont {Fogler},\ and\ \citenamefont {Garc{\'\i}a~de
  Abajo}}]{Basov2016}%
  \BibitemOpen
  \bibfield  {author} {\bibinfo {author} {\bibfnamefont {D.~N.}\ \bibnamefont
  {Basov}}, \bibinfo {author} {\bibfnamefont {M.~M.}\ \bibnamefont {Fogler}}, \
  and\ \bibinfo {author} {\bibfnamefont {F.~J.}\ \bibnamefont {Garc{\'\i}a~de
  Abajo}},\ }\bibfield  {title} {\enquote {\bibinfo {title} {Polaritons in van
  der {W}aals materials},}\ }\href {\doibase 10.1126/science.aag1992}
  {\bibfield  {journal} {\bibinfo  {journal} {Science}\ }\textbf {\bibinfo
  {volume} {354}} (\bibinfo {year} {2016}),\
  10.1126/science.aag1992}\BibitemShut {NoStop}%
\bibitem [{\citenamefont {Savona}\ \emph {et~al.}(1995)\citenamefont {Savona},
  \citenamefont {Andreani}, \citenamefont {Schwendimann},\ and\ \citenamefont
  {Quattropani}}]{savona1995quantum}%
  \BibitemOpen
  \bibfield  {author} {\bibinfo {author} {\bibfnamefont {Vincenzo}\
  \bibnamefont {Savona}}, \bibinfo {author} {\bibfnamefont {LC}~\bibnamefont
  {Andreani}}, \bibinfo {author} {\bibfnamefont {P}~\bibnamefont
  {Schwendimann}}, \ and\ \bibinfo {author} {\bibfnamefont {A}~\bibnamefont
  {Quattropani}},\ }\bibfield  {title} {\enquote {\bibinfo {title} {Quantum
  well excitons in semiconductor microcavities: unified treatment of weak and
  strong coupling regimes},}\ }\href@noop {} {\bibfield  {journal} {\bibinfo
  {journal} {Sol. St. Commun.}\ }\textbf {\bibinfo {volume} {93}},\ \bibinfo
  {pages} {733--739} (\bibinfo {year} {1995})}\BibitemShut {NoStop}%
\bibitem [{\citenamefont {Plechinger}\ \emph {et~al.}(2012)\citenamefont
  {Plechinger}, \citenamefont {Schrettenbrunner}, \citenamefont {Eroms},
  \citenamefont {Weiss}, \citenamefont {Schueller},\ and\ \citenamefont
  {Korn}}]{plechinger2012low}%
  \BibitemOpen
  \bibfield  {author} {\bibinfo {author} {\bibfnamefont {Gerd}\ \bibnamefont
  {Plechinger}}, \bibinfo {author} {\bibfnamefont {F-X}\ \bibnamefont
  {Schrettenbrunner}}, \bibinfo {author} {\bibfnamefont {Jonathan}\
  \bibnamefont {Eroms}}, \bibinfo {author} {\bibfnamefont {Dieter}\
  \bibnamefont {Weiss}}, \bibinfo {author} {\bibfnamefont {Christian}\
  \bibnamefont {Schueller}}, \ and\ \bibinfo {author} {\bibfnamefont {Tobias}\
  \bibnamefont {Korn}},\ }\bibfield  {title} {\enquote {\bibinfo {title}
  {Low-temperature photoluminescence of oxide-covered single-layer
  {M}o{S}$_2$},}\ }\href@noop {} {\bibfield  {journal} {\bibinfo  {journal}
  {Phys. Stat. Sol.}\ }\textbf {\bibinfo {volume} {6}},\ \bibinfo {pages}
  {126--128} (\bibinfo {year} {2012})}\BibitemShut {NoStop}%
\bibitem [{\citenamefont {Tongay}\ \emph {et~al.}(2013)\citenamefont {Tongay},
  \citenamefont {Suh}, \citenamefont {Ataca}, \citenamefont {Fan},
  \citenamefont {Luce}, \citenamefont {Kang}, \citenamefont {Liu},
  \citenamefont {Ko}, \citenamefont {Raghunathanan}, \citenamefont {Zhou},
  \citenamefont {Ogletree}, \citenamefont {Li}, \citenamefont {Grossman},\ and\
  \citenamefont {Wu}}]{tongay2013defects}%
  \BibitemOpen
  \bibfield  {author} {\bibinfo {author} {\bibfnamefont {Sefaattin}\
  \bibnamefont {Tongay}}, \bibinfo {author} {\bibfnamefont {Joonki}\
  \bibnamefont {Suh}}, \bibinfo {author} {\bibfnamefont {Can}\ \bibnamefont
  {Ataca}}, \bibinfo {author} {\bibfnamefont {Wen}\ \bibnamefont {Fan}},
  \bibinfo {author} {\bibfnamefont {Alexander}\ \bibnamefont {Luce}}, \bibinfo
  {author} {\bibfnamefont {Jeong~Seuk}\ \bibnamefont {Kang}}, \bibinfo {author}
  {\bibfnamefont {Jonathan}\ \bibnamefont {Liu}}, \bibinfo {author}
  {\bibfnamefont {Changhyun}\ \bibnamefont {Ko}}, \bibinfo {author}
  {\bibfnamefont {Rajamani}\ \bibnamefont {Raghunathanan}}, \bibinfo {author}
  {\bibfnamefont {Jian}\ \bibnamefont {Zhou}}, \bibinfo {author} {\bibfnamefont
  {Frank}\ \bibnamefont {Ogletree}}, \bibinfo {author} {\bibfnamefont {Jingbo}\
  \bibnamefont {Li}}, \bibinfo {author} {\bibfnamefont {Jeffrey~C.}\
  \bibnamefont {Grossman}}, \ and\ \bibinfo {author} {\bibfnamefont {Junqiao}\
  \bibnamefont {Wu}},\ }\bibfield  {title} {\enquote {\bibinfo {title} {Defects
  activated photoluminescence in two-dimensional semiconductors: interplay
  between bound, charged, and free excitons},}\ }\href@noop {} {\bibfield
  {journal} {\bibinfo  {journal} {Sci. Rep.}\ }\textbf {\bibinfo {volume} {3}}
  (\bibinfo {year} {2013})}\BibitemShut {NoStop}%
\bibitem [{\citenamefont {Laussy}\ \emph {et~al.}(2009)\citenamefont {Laussy},
  \citenamefont {Del~Valle},\ and\ \citenamefont
  {Tejedor}}]{laussy2009luminescence}%
  \BibitemOpen
  \bibfield  {author} {\bibinfo {author} {\bibfnamefont {Fabrice~P}\
  \bibnamefont {Laussy}}, \bibinfo {author} {\bibfnamefont {Elena}\
  \bibnamefont {Del~Valle}}, \ and\ \bibinfo {author} {\bibfnamefont {Carlos}\
  \bibnamefont {Tejedor}},\ }\bibfield  {title} {\enquote {\bibinfo {title}
  {Luminescence spectra of quantum dots in microcavities. {I}. {B}osons},}\
  }\href@noop {} {\bibfield  {journal} {\bibinfo  {journal} {Phys. Rev. B}\
  }\textbf {\bibinfo {volume} {79}},\ \bibinfo {pages} {235325} (\bibinfo
  {year} {2009})}\BibitemShut {NoStop}%
\bibitem [{\citenamefont {Moody}\ \emph {et~al.}(2015)\citenamefont {Moody},
  \citenamefont {Dass}, \citenamefont {Hao}, \citenamefont {Chen},
  \citenamefont {Li}, \citenamefont {Singh}, \citenamefont {Tran},
  \citenamefont {Clark}, \citenamefont {Xu}, \citenamefont {Bergh{\"a}user},
  \citenamefont {Malic}, \citenamefont {Knorr},\ and\ \citenamefont
  {Li}}]{Moody2015}%
  \BibitemOpen
  \bibfield  {author} {\bibinfo {author} {\bibfnamefont {Galan}\ \bibnamefont
  {Moody}}, \bibinfo {author} {\bibfnamefont {Chandriker~Kavir}\ \bibnamefont
  {Dass}}, \bibinfo {author} {\bibfnamefont {Kai}\ \bibnamefont {Hao}},
  \bibinfo {author} {\bibfnamefont {Chang-Hsiao}\ \bibnamefont {Chen}},
  \bibinfo {author} {\bibfnamefont {Lain-Jong}\ \bibnamefont {Li}}, \bibinfo
  {author} {\bibfnamefont {Akshay}\ \bibnamefont {Singh}}, \bibinfo {author}
  {\bibfnamefont {Kha}\ \bibnamefont {Tran}}, \bibinfo {author} {\bibfnamefont
  {Genevieve}\ \bibnamefont {Clark}}, \bibinfo {author} {\bibfnamefont
  {Xiaodong}\ \bibnamefont {Xu}}, \bibinfo {author} {\bibfnamefont {Gunnar}\
  \bibnamefont {Bergh{\"a}user}}, \bibinfo {author} {\bibfnamefont {Ermin}\
  \bibnamefont {Malic}}, \bibinfo {author} {\bibfnamefont {Andreas}\
  \bibnamefont {Knorr}}, \ and\ \bibinfo {author} {\bibfnamefont {Xiaoqin}\
  \bibnamefont {Li}},\ }\bibfield  {title} {\enquote {\bibinfo {title}
  {Intrinsic homogeneous linewidth and broadening mechanisms of excitons in
  monolayer transition metal dichalcogenides},}\ }\href
  {http://dx.doi.org/10.1038/ncomms9315} {\bibfield  {journal} {\bibinfo
  {journal} {Nature Commun.}\ }\textbf {\bibinfo {volume} {6}},\ \bibinfo
  {pages} {8315} (\bibinfo {year} {2015})}\BibitemShut {NoStop}%
\bibitem [{\citenamefont {Robert}\ \emph {et~al.}(2016)\citenamefont {Robert},
  \citenamefont {Lagarde}, \citenamefont {Cadiz}, \citenamefont {Wang},
  \citenamefont {Lassagne}, \citenamefont {Amand}, \citenamefont {Balocchi},
  \citenamefont {Renucci}, \citenamefont {Tongay}, \citenamefont {Urbaszek},\
  and\ \citenamefont {Marie}}]{Robert2016}%
  \BibitemOpen
  \bibfield  {author} {\bibinfo {author} {\bibfnamefont {C.}~\bibnamefont
  {Robert}}, \bibinfo {author} {\bibfnamefont {D.}~\bibnamefont {Lagarde}},
  \bibinfo {author} {\bibfnamefont {F.}~\bibnamefont {Cadiz}}, \bibinfo
  {author} {\bibfnamefont {G.}~\bibnamefont {Wang}}, \bibinfo {author}
  {\bibfnamefont {B.}~\bibnamefont {Lassagne}}, \bibinfo {author}
  {\bibfnamefont {T.}~\bibnamefont {Amand}}, \bibinfo {author} {\bibfnamefont
  {A.}~\bibnamefont {Balocchi}}, \bibinfo {author} {\bibfnamefont
  {P.}~\bibnamefont {Renucci}}, \bibinfo {author} {\bibfnamefont
  {S.}~\bibnamefont {Tongay}}, \bibinfo {author} {\bibfnamefont
  {B.}~\bibnamefont {Urbaszek}}, \ and\ \bibinfo {author} {\bibfnamefont
  {X.}~\bibnamefont {Marie}},\ }\bibfield  {title} {\enquote {\bibinfo {title}
  {Exciton radiative lifetime in transition metal dichalcogenide monolayers},}\
  }\href {\doibase 10.1103/PhysRevB.93.205423} {\bibfield  {journal} {\bibinfo
  {journal} {Phys. Rev. B}\ }\textbf {\bibinfo {volume} {93}},\ \bibinfo
  {pages} {205423} (\bibinfo {year} {2016})}\BibitemShut {NoStop}%
\bibitem [{\citenamefont {Lagarde}\ \emph {et~al.}(2014)\citenamefont
  {Lagarde}, \citenamefont {Bouet}, \citenamefont {Marie}, \citenamefont {Zhu},
  \citenamefont {Liu}, \citenamefont {Amand}, \citenamefont {Tan},\ and\
  \citenamefont {Urbaszek}}]{lagarde2014carrier}%
  \BibitemOpen
  \bibfield  {author} {\bibinfo {author} {\bibfnamefont {D}~\bibnamefont
  {Lagarde}}, \bibinfo {author} {\bibfnamefont {L}~\bibnamefont {Bouet}},
  \bibinfo {author} {\bibfnamefont {X}~\bibnamefont {Marie}}, \bibinfo {author}
  {\bibfnamefont {CR}~\bibnamefont {Zhu}}, \bibinfo {author} {\bibfnamefont
  {BL}~\bibnamefont {Liu}}, \bibinfo {author} {\bibfnamefont {T}~\bibnamefont
  {Amand}}, \bibinfo {author} {\bibfnamefont {PH}~\bibnamefont {Tan}}, \ and\
  \bibinfo {author} {\bibfnamefont {B}~\bibnamefont {Urbaszek}},\ }\bibfield
  {title} {\enquote {\bibinfo {title} {Carrier and polarization dynamics in
  monolayer mos 2},}\ }\href@noop {} {\bibfield  {journal} {\bibinfo  {journal}
  {Phys. Rev. Lett.}\ }\textbf {\bibinfo {volume} {112}},\ \bibinfo {pages}
  {047401} (\bibinfo {year} {2014})}\BibitemShut {NoStop}%
\bibitem [{\citenamefont {Palummo}\ \emph {et~al.}(2015)\citenamefont
  {Palummo}, \citenamefont {Bernardi},\ and\ \citenamefont
  {Grossman}}]{Palummo2015}%
  \BibitemOpen
  \bibfield  {author} {\bibinfo {author} {\bibfnamefont {Maurizia}\
  \bibnamefont {Palummo}}, \bibinfo {author} {\bibfnamefont {Marco}\
  \bibnamefont {Bernardi}}, \ and\ \bibinfo {author} {\bibfnamefont
  {Jeffrey~C.}\ \bibnamefont {Grossman}},\ }\bibfield  {title} {\enquote
  {\bibinfo {title} {Exciton radiative lifetimes in two-dimensional transition
  metal dichalcogenides},}\ }\href {\doibase 10.1021/nl503799t} {\bibfield
  {journal} {\bibinfo  {journal} {Nano Letters}\ }\textbf {\bibinfo {volume}
  {15}},\ \bibinfo {pages} {2794--2800} (\bibinfo {year} {2015})}\BibitemShut
  {NoStop}%
\bibitem [{\citenamefont {Lee}\ \emph {et~al.}(2012)\citenamefont {Lee},
  \citenamefont {Zhang}, \citenamefont {Zhang}, \citenamefont {Chang},
  \citenamefont {Lin}, \citenamefont {Chang}, \citenamefont {Yu}, \citenamefont
  {Wang}, \citenamefont {Chang}, \citenamefont {Li},\ and\ \citenamefont
  {Lin}}]{lee2012synthesis}%
  \BibitemOpen
  \bibfield  {author} {\bibinfo {author} {\bibfnamefont {Yi-Hsien}\
  \bibnamefont {Lee}}, \bibinfo {author} {\bibfnamefont {Xin-Quan}\
  \bibnamefont {Zhang}}, \bibinfo {author} {\bibfnamefont {Wenjing}\
  \bibnamefont {Zhang}}, \bibinfo {author} {\bibfnamefont {Mu-Tung}\
  \bibnamefont {Chang}}, \bibinfo {author} {\bibfnamefont {Cheng-Te}\
  \bibnamefont {Lin}}, \bibinfo {author} {\bibfnamefont {Kai-Di}\ \bibnamefont
  {Chang}}, \bibinfo {author} {\bibfnamefont {Ya-Chu}\ \bibnamefont {Yu}},
  \bibinfo {author} {\bibfnamefont {Jacob Tse-Wei}\ \bibnamefont {Wang}},
  \bibinfo {author} {\bibfnamefont {Chia-Seng}\ \bibnamefont {Chang}}, \bibinfo
  {author} {\bibfnamefont {Lain-Jong}\ \bibnamefont {Li}}, \ and\ \bibinfo
  {author} {\bibfnamefont {Tsung-Wu}\ \bibnamefont {Lin}},\ }\bibfield  {title}
  {\enquote {\bibinfo {title} {Synthesis of large-area {M}o{S}$_2$ atomic
  layers with chemical vapor deposition},}\ }\href@noop {} {\bibfield
  {journal} {\bibinfo  {journal} {Adv. Mater.}\ }\textbf {\bibinfo {volume}
  {24}},\ \bibinfo {pages} {2320--2325} (\bibinfo {year} {2012})}\BibitemShut
  {NoStop}%
\bibitem [{\citenamefont {Lin}\ \emph {et~al.}(2011)\citenamefont {Lin},
  \citenamefont {Jin}, \citenamefont {Lee}, \citenamefont {Jen}, \citenamefont
  {Suenaga},\ and\ \citenamefont {Chiu}}]{lin2011clean}%
  \BibitemOpen
  \bibfield  {author} {\bibinfo {author} {\bibfnamefont {Yung-Chang}\
  \bibnamefont {Lin}}, \bibinfo {author} {\bibfnamefont {Chuanhong}\
  \bibnamefont {Jin}}, \bibinfo {author} {\bibfnamefont {Jung-Chi}\
  \bibnamefont {Lee}}, \bibinfo {author} {\bibfnamefont {Shou-Feng}\
  \bibnamefont {Jen}}, \bibinfo {author} {\bibfnamefont {Kazu}\ \bibnamefont
  {Suenaga}}, \ and\ \bibinfo {author} {\bibfnamefont {Po-Wen}\ \bibnamefont
  {Chiu}},\ }\bibfield  {title} {\enquote {\bibinfo {title} {Clean transfer of
  graphene for isolation and suspension},}\ }\href@noop {} {\bibfield
  {journal} {\bibinfo  {journal} {ACS Nano}\ }\textbf {\bibinfo {volume} {5}},\
  \bibinfo {pages} {2362--2368} (\bibinfo {year} {2011})}\BibitemShut {NoStop}%
\bibitem [{\citenamefont {Park}\ \emph {et~al.}(2010)\citenamefont {Park},
  \citenamefont {Meyer}, \citenamefont {Roth},\ and\ \citenamefont
  {Sk{\'a}kalov{\'a}}}]{park2010growth}%
  \BibitemOpen
  \bibfield  {author} {\bibinfo {author} {\bibfnamefont {Hye~Jin}\ \bibnamefont
  {Park}}, \bibinfo {author} {\bibfnamefont {Jannik}\ \bibnamefont {Meyer}},
  \bibinfo {author} {\bibfnamefont {Siegmar}\ \bibnamefont {Roth}}, \ and\
  \bibinfo {author} {\bibfnamefont {Viera}\ \bibnamefont {Sk{\'a}kalov{\'a}}},\
  }\bibfield  {title} {\enquote {\bibinfo {title} {Growth and properties of
  few-layer graphene prepared by chemical vapor deposition},}\ }\href@noop {}
  {\bibfield  {journal} {\bibinfo  {journal} {Carbon}\ }\textbf {\bibinfo
  {volume} {48}},\ \bibinfo {pages} {1088--1094} (\bibinfo {year}
  {2010})}\BibitemShut {NoStop}%
\end{thebibliography}%

\end{document}